\documentclass[a4paper,11pt]{article}
\usepackage{jcappub} 

\makeatletter
\gdef\@fpheader{ }
\makeatother

\newcommand{\N}{{\mathcal{N}}}
\newcommand{\phie}{{\phi_e}} 
\title{Quantum Diffusion in Sharp Transition to Non-Slow-Roll Phase}

\author[a]{Nahid Ahmadi}
\author[a,b]{Mahdiyar Noorbala}
\author[a]{Niloufar Feyzabadi}
\author[a]{Fatemeh Eghbalpoor}
\author[a]{Zahra Ahmadi}

\affiliation[a]{Department of Physics, University of Tehran, Iran, P.O.~Box 14395-547}
\affiliation[b]{School of Astronomy, Institute for Research in Fundamental Sciences (IPM), P.O.~Box 19395-5531, Tehran, Iran}

\emailAdd{nahmadi@ut.ac.ir}
\emailAdd{mnoorbala@ut.ac.ir}
\emailAdd{niloufar\_feyz@ut.ac.ir}
\emailAdd{f.eghbalpoor90@alumni.ut.ac.ir}
\emailAdd{zr.ahmadi@ut.ac.ir}

\begin{document}

\abstract{Transitions between different inflationary slow-roll scenarios are known to provide short non-slow-roll periods with non-trivial consequences.  We consider the effect of quantum diffusion on the inflationary dynamics in a transition process. Using the stochastic $\delta\mathcal{N}$ formalism, we follow the detailed evolution of noises through a sharp transition modeled by the Starobinsky potential, although some of our results apply to any sharp transition.  We find how the stochastic noise induced by the transition affects the coarse-grained fields.  We then consider the special case that the potential is flat after the transition.  It is found that the particular noise we obtain cannot drive the inflaton past the classically unreachable field values.  By deriving the characteristic function, we also study the tail behavior for the distribution of curvature perturbations $\zeta$, which we find to decay faster than $\exp(-3\zeta)$.}

\maketitle

\section{Introduction}\label{sec:introduction}


The inflationary scenario is the leading paradigm for describing the early universe~\cite{Starobinsky:1980te, Sato:1980yn, Guth:1980zm, Linde:1981mu, Albrecht:1982wi, Linde:1983gd, Mukhanov:1981xt, Hawking:1982cz,  Starobinsky:1982ee, Guth:1982ec, Bardeen:1983qw, Starobinsky:1979ty}. Although the slow-roll inflationary models yield a simple realization of inflation with viable experimental predictions especially on CMB scales, it is possible to violate the slow-roll assumption at field values corresponding to smaller scales where the models are not constrained by data. This is often done at the last stages of inflation by introduction of small and short bursts of deviation from slow-roll wherein the first slow-roll parameter $\varepsilon_1$ falls exponentially fast~\cite{Tsamis:2003px, Kinney:2005vj, Namjoo:2012aa}. In a transition to a non-slow-roll phase, one has a growing superhorizon mode (in addition to the usual constant mode), which is well studied in standard perturbation theory. The impact on non-Gaussianity of the reverse transitions to slow-roll was also analyzed in Ref.~\cite{Cai:2018dkf}.
In fact, the customized inflationary potentials exploit this decrease in $\varepsilon_1$ to produce peaks in power spectrum which are seven orders of magnitude above the amplitudes observed in CMB scales and sufficient for ultra-compact objects such as Primordial Black Holes (PBHs) to form.  However, the power spectrum is at best a perturbative characterization of the information about the tail of the probability distribution of curvature perturbations. Due to ultraviolet sensitivity of inflation and unlike the CMB scales, the treatment of these large perturbations require a non-perturbative approach to take into account the back-reaction of the quantum diffusion induced by small scales.

A non-perturbative description of coarse-grained fields, which is consistent with quantum field theoretic techniques, is given by the stochastic inflation formalism~\cite{Vilenkin:1983xp, Vilenkin:1983xq, Linde:1986fd, Starobinsky:1986fx, Rey:1986zk, Aryal:1987vn, Sasaki:1987gy, Nambu:1987ef, Nambu:1988je, Kandrup:1988sc, Nakao:1988yi, Nambu:1989uf, Mollerach:1990zf, Linde:1993xx, Linde:1996hg, Starobinsky:1994bd, Kunze:2006tu, Prokopec:2007ak, Prokopec:2008gw, Tsamis:2005hd, Enqvist:2008kt, Finelli:2008zg, Finelli:2010sh, Garbrecht:2013coa, Garbrecht:2014dca, Burgess:2014eoa, Burgess:2015ajz, Boyanovsky:2015tba, Boyanovsky:2015jen, Fujita:2017lfu, Gorbenko:2019rza, Mirbabayi:2019qtx, Mirbabayi:2020vyt, Cohen:2021fzf}. This formalism deals with the super Hubble (classical) part of quantum fields driving the inflation in a stochastic way.  In fact those are sourced by the small scale fluctuations, as a classical noise, and follow Langevin equations. In practice, this approach is combined with the $\delta N$ formalism~\cite{Sasaki:1995aw, Sasaki:1998ug, Lyth:2004gb, Wands:2000dp, Lyth:2005fi, Talebian-Ashkezari:2016llx} to give birth to stochastic $\delta \cal N$ approach~\cite{Fujita:2013cna, Fujita:2014tja, Vennin:2015hra, Vennin:2016wnk, Assadullahi:2016gkk, Grain:2017dqa, Pattison:2018bct, Noorbala:2018zlv, Firouzjahi:2018vet, Noorbala:2019kdd, Pattison:2019hef, Talebian:2019opf, Firouzjahi:2020jrj, Ando:2020fjm, Talebian:2022jkb}. Relying on the separate universe approach, it provides a framework in which the statistics of coarse-grained cosmological perturbations are calculated. The probability distribution (PDF) of curvature perturbations is identified with the distribution of first passage time through the end of inflation, starting from some initial conditions.
 
The stochastic $\delta \cal N$ approach has been notably used for treating the regimes that part of the velocity of the inflaton, which is induced by the potential gradient, is small; so the quantum diffusion effects dominate. This is typically the case in flat potentials in which the inflation proceeds in the so-called ultra-slow-roll (USR) regime.  Since there is no gradient-induced velocity, the Hubble friction restricts the field excursion to a finite value, which we call the classically accessible field values.  These non-attractor phases typically last for a few $e$-folds before approaching a new attractor. In any case they must be terminated to allow for inflation to end, therefore a transient non-slow-roll phase is  primarily confined between two slow-roll phases. If the transition to the USR phase is sharp enough, then it affects the stochastic noise~\cite{Ballesteros:2020sre}.  In the earlier researches  using the stochastic $\delta \cal N$ formalism, the focus was on the USR phase alone and the transition was not taken into account; so the nearly constant noise amplitude $H/2\pi$ was simply used throughout the evolution~\cite{Pattison:2021oen}.  Based on this form of the noise, it is concluded in Ref.~\cite{Pattison:2021oen} that diffusion assists the inflaton on a flat potential to reach beyond the classically accessible field values.

In this work, we revisit the non-perturbative evolution of the large cosmological perturbations for potentials whose slope undergoes a rapid change over a tiny interval of field values.  As a concrete example, we present our results for the Starobinsky potential~\cite{Starobinsky:1992ts} in which a sudden transition changes the potential slope.  We focus on an instant transition phase which is characterized by a subsequent relaxation period of time for phase space variables as well as the slow roll parameters to reach their final attractor values.  We elaborate on the effect of this transition on the noise amplitude.  In particular, we find that the velocity noise has a quick rise after the transition.  In a subsequent stage, all noise amplitudes decay at the same rate as the background field velocity.  As a consequence, the inflaton can no longer reach the classically inaccessible field values, even with the help of diffusion.  We find the stochastic effect of a transient non-slow-roll phase on inflationary observables including the power spectrum.


A universal feature of the PDFs, that arises when quantum diffusion is incorporated, and makes them different from Gaussian distributions, is the heavy exponential tail~\cite{Ezquiaga:2019ftu}.  In regimes with very flat inflationary potential, when the gradient-induced velocity cannot proceed the inflation, quantum diffusion plays role in determining the PBH abundance. Since PBHs form in regimes with large curvature perturbations, beyond a threshold, this has been an early motivation for studying the inflection points, flat potentials and analyzing the PDF tail~\cite{Biagetti:2018pjj, Pattison:2017mbe, Panagopoulos:2019ail, Ballesteros:2020sre, Celoria:2021vjw, Pattison:2021oen, Figueroa:2020jkf, Figueroa:2021zah, Cohen:2021jbo, Hooshangi:2021ubn, Cai:2021zsp, Hooshangi:2022lao}.  In this paper we study the effect of the transition-induced form of noise amplitudes on the tail, compute the decay exponent of the PDF in our model and calculate its dependence on the diffusion strength. 


The rest of this paper is organized as follows.  In Section~\ref{sec:background} we review the Starobinsky model at the background and perturbative level.  Next we give a brief review of the stochastic $\delta\cal N$ formalism in Section~\ref{sec:review-stochastic}, before calculating the noise amplitudes in the Starobinsky model in Section~\ref{sec:noise}.  Then we employ the stochastic $\delta\cal N$ techniques in Section~\ref{sec:perturbative} to perturbatively calculate the statistical properties of the coarse-grained field after the transition at the end of the relaxation period.  A non-perturbative treatment is presented in Section~\ref{sec:nonperturbative} for the special case of a transition to USR inflation.  Finally we summarize our results and conclude in Section~\ref{sec:conclusions}.


\section{The Starobinsky Model}\label{sec:background}

In this section, we present a detailed analysis of the Starobinsky model as an inflationary model with transition. We discuss the background dynamics as well as the comoving curvature perturbation induced by the inflationary field that experiences the transition, and its power spectrum in standard perturbation theory. 

\subsection{Background Dynamics}

In this section we review the dynamics of a scenario with a sharp transition in the inflationary potential.  As a prototype we work with the Starobinsky potential~\cite{Starobinsky:1992ts} which consists of two linear pieces characterized by dimensionless parameters $\alpha,\beta$:
\begin{equation}\label{V}
V\left(\phi\right) = 
\begin{cases}
V_{0} \left[1+\alpha\left(\phi-\phi_{0}\right)\right], & \phi>\phi_{0}, \\
V_{0} \left[1+\beta\left(\phi-\phi_{0}\right)\right], & \phi<\phi_{0}.
\end{cases}
\end{equation}
We concentrate on the $\alpha>\beta>0$ case.  In this model, inflation starts from a slow-roll phase $ \phi>\phi_{0} $ and when the inflaton field crosses the transition point $\phi=\phi_{0}$, a phase of USR inflation starts \cite{Namjoo:2012aa}. This phase is unstable and after a while the system reenters the slow-roll phase.  

The first two slow-roll parameters are
\begin{equation}
\varepsilon_1 = -\frac{\dot H}{H^2} = \frac12 \pi^2, \qquad
\varepsilon_2 = \frac{d}{dN}\log\varepsilon_1 = \frac2\pi \frac{d\pi}{dN},
\end{equation}
where $N$ is the number of $e$-folds and $\pi \equiv d\phi/dN$.\footnote{Note that the canonical conjugate of $\phi$ is $Ha^3\pi$, not $\pi$.}  The classical equations of motion in an FLRW cosmology are given by the Friedman and Klein-Gordon equations\footnote{We work in the reduced Planck mass units throughout: $M_{\rm Pl}=1/\sqrt{8\pi G_{\rm N}}=1$.}
\begin{equation}\label{Friedmann}
3H^2 = \frac{1}{2} H^2 \left(\frac{d\phi}{dN}\right)^2 + V,
\end{equation}
\begin{equation}\label{KG}
H^{2}\frac{d^{2}\phi}{dN^2}+\left(3-\varepsilon_{1}\right)H^{2}\frac{d\phi}{dN}+\frac{dV}{d\phi}=0.
\end{equation}

\begin{figure}
\centering
\includegraphics[scale=.5]{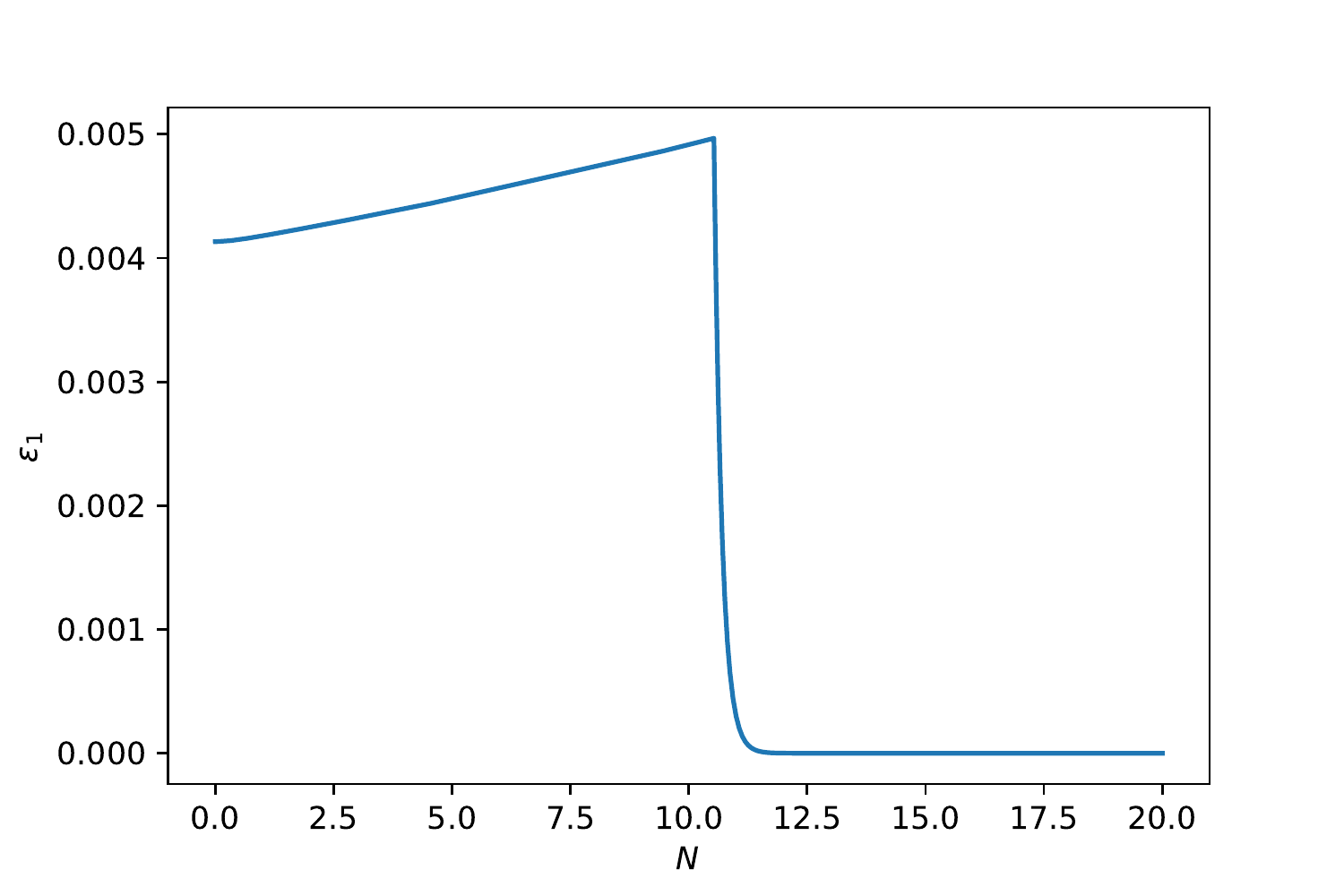}
\includegraphics[scale=.5]{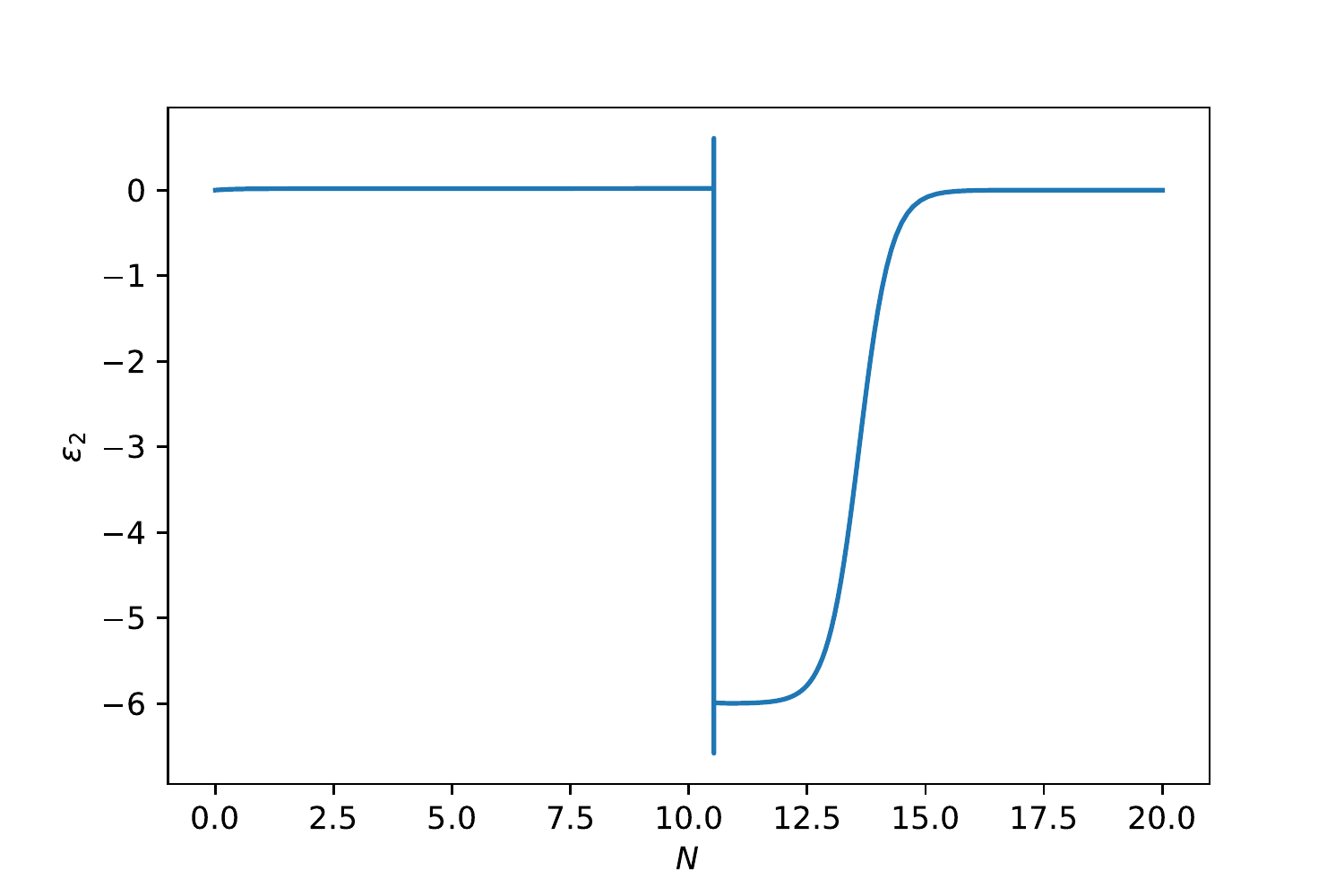}
\caption{The slow-roll parameters $\varepsilon_1$ and $\varepsilon_2$ for the parameters $\phi_0=0$, $V_0=10^{-10}M_{\rm Pl}^4$, $\alpha=0.1$, and $\beta=10^{-5}$.}
\label{fig: slow-roll parameters}
\end{figure}

\begin{figure}
\includegraphics[scale=.5]{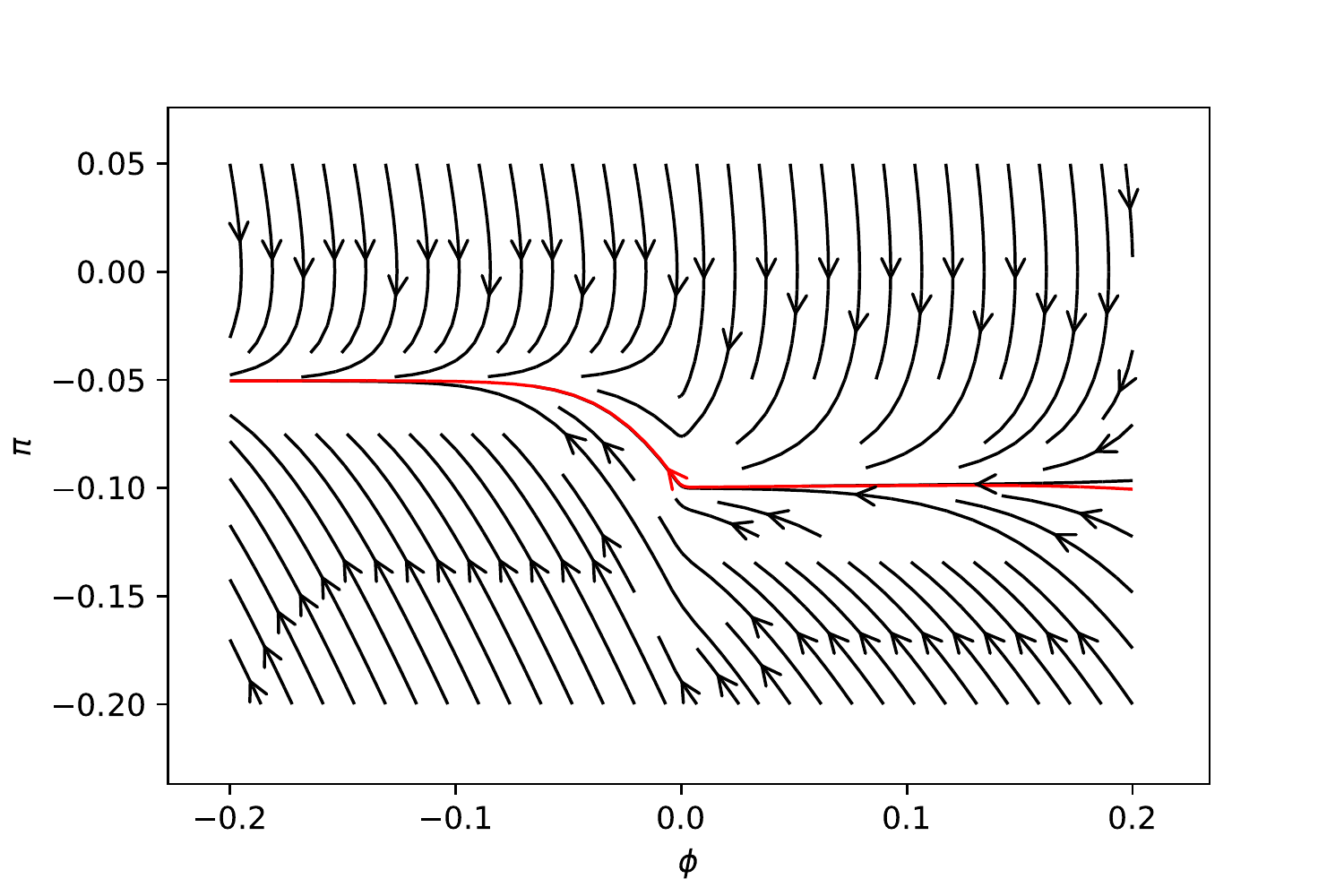}
\includegraphics[scale=.5]{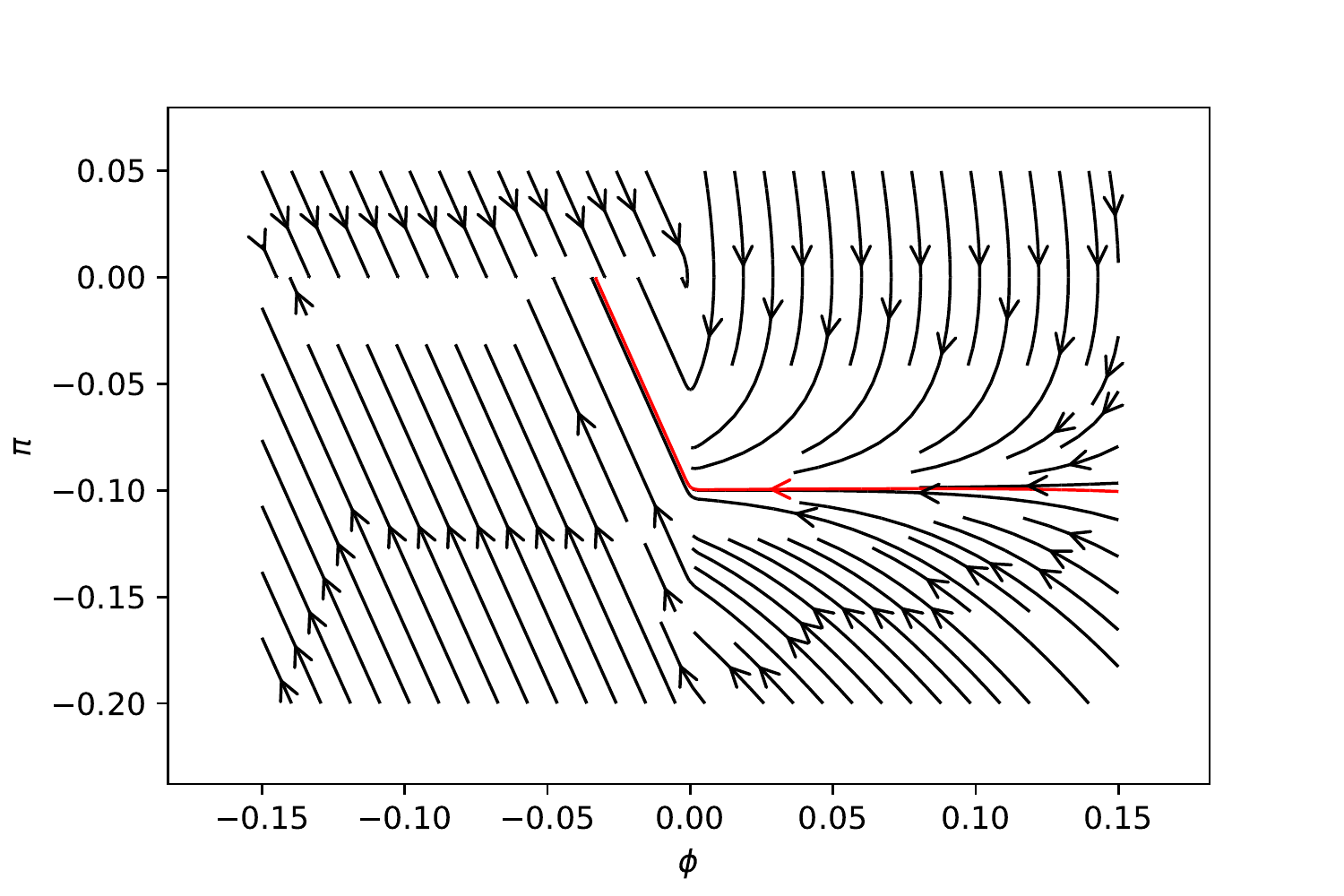}
\caption{Background trajectories on the phase space for the Starobinsky potential~\eqref{V} with parameters $\phi_0=0$, $V_0=10^{-10}M_{\rm Pl}^4$, $\alpha=0.1$, and with $\beta=0.05$ (left), $\beta=0$ (right).  The curve highlighted with red represents the background trajectory that we consider throughout the paper.}\label{fig:phase}
\end{figure}

Let us follow the evolution of $\phi$ in different phases. In the first slow-roll phase, the field rolls down the potential and reaches the attractor, where $\varepsilon_1, \varepsilon_2 \ll 1$.  So writing Eqs.~\eqref{Friedmann} and \eqref{KG} in the equivalent form
\begin{equation}
\left( 3 - \varepsilon_{1} + \frac{\varepsilon_2}{2} \right) \pi + \left( 3 - \varepsilon_{1} \right) \frac{1}{V}\frac{dV}{d\phi} = 0,
\end{equation}
we find $\pi+V_{,\phi}/V=0$ which gives the equation of the attractor trajectory in the phase space
\begin{equation}
\pi + \frac{\alpha}{1+\alpha\left(\phi-\phi_{0}\right)} = 0.
\end{equation}
As $\phi$ approaches $\phi_{0}$, $\pi$ slightly decreases toward $-\alpha$ as can be seen in Figures.~\ref{fig: slow-roll parameters} and \ref{fig:phase}.  We require that $\alpha\ll1$ in order that $\varepsilon_{1} \to \alpha^2/2 \ll 1$; similarly $\varepsilon_2 \to 2\alpha^2 \ll 1$.  
The expressions for $\phi$ and $\pi$ simplify to 
\begin{equation}\label{SR-background}
\phi \approx \phi_0 + \pi_0 n, \qquad \pi \approx \pi_0 + \alpha^2(\phi-\phi_0).
\end{equation}
Here $n=N-N_{0}$ and $N_{0}$ is the $e$-fold number wherein the field crosses $\phi=\phi_{0}$ and the USR phase starts.  The terminal velocity $\pi_0 \equiv -\alpha$ of the first slow-roll phase will be inherited as the initial velocity of the post-transition phase.  

Next consider the post-transition phase.  At sufficiently late times after the transition, when slow-roll has been restored, similar relations like above, with $\alpha$ replaced by $\beta$, are expected.  But immediately after the transition, we have a non-slow-roll phase.  The quasi de~Sitter approximation $\varepsilon_{1}\ll1$ is valid even though the slow-roll approximation breaks down temporarily after $\phi_{0}$. Under this assumption, Eq.~\eqref{KG} becomes
\begin{equation}\label{KG-dS}
\frac{d^2\phi}{dN^2} + 3\frac{d\phi}{dN} + 3\beta=0,
\end{equation}
whose solution is
\begin{equation}\label{USR-background}
\begin{aligned}
\phi\left(N\right)&=\phi_{0}-\frac{\beta}{3}D_{0}\left[1-e^{-3n}\right]-\beta n, \\
\pi\left(N\right)&= -\beta\left[1+D_{0}e^{-3n}\right].
\end{aligned} 
\end{equation}
We have defined a dimensionless function
\begin{equation}\label{def:D}
D(\pi)\equiv-\left(1+\frac{\pi}{\beta}\right),
\end{equation}
whose value at $N=N_{0}$ is denoted by $D_{0}$. This is easily seen to satisfy $D = D_{0} e^{-3n}$, which implies that $\pi\to-\beta$ exponentially at late times (c.f.\ $\pi\to-\alpha$, which occurs linearly in $n$ at the end of the first slow-roll phase). 
As can be seen in Figure~\ref{fig: slow-roll parameters}, for $\alpha\gg\beta$, the second slow-roll parameter $\varepsilon_{2}$ quickly approaches $-6$ immediately after the transition and then exhibits a sharp rise a few $e$-folds after $N_0$. 
This can also be seen analytically from
\begin{equation}
\varepsilon_{2} = \frac{-6D_0e^{-3n}}{1+D_0e^{-3n}},
\end{equation}
by noting that $D_0 = -1 + \alpha/\beta \approx \alpha/\beta \gg 1$.  Half way between the USR value $\varepsilon_2=-6$ and the slow-roll limit $\varepsilon_2=0$, the value of the second slow-roll parameter is $\varepsilon_2=-3$ which is attained when $D_0e^{-3n}=1$.  We may therefore characterize $n=\frac13\log D_0$ as the end of the USR phase.

Put together, the preceding remarks clarify
the response of the scalar field to the slope change in potential. This may be checked in background trajectories on the phase space, displayed in Figure~\ref{fig:phase}. In all the trajectories, the force due to the potential dominates and the classical trajectory approaches the slow-roll attractor. In the $\beta=0$ case, the classical velocity vanishes asymptotically and the field approaches $\phi_{\rm min} = \phi_{0}+\pi_{0}/3$ as $N\to\infty$. One therefore expects, in this limiting case, that the stochastic effects can potentially play an important role in the absence of the classical drift. Before we scrutinize these stochastic effects, let us discuss the perturbations and the power spectrum in this model.
\subsection{Perturbations}\label{perturbations}


We now come to the comoving metric perturbation $\cal R$ which is related to the field perturbation $\delta\phi$ in the spatially flat gauge via ${\cal R} = H\delta\phi/\dot\phi$.
By quantizing the curvature perturbation, we get
\begin{equation}
\hat{\mathcal{R}}\left(\eta,\textbf{x}\right)=\int \frac{d^3k}{\left(2\pi\right)^{\frac{3}{2}}}\left[\hat{a}_{\textbf{k}}{\cal R}_{k}\left(\eta\right)e^{i\textbf{k}\cdot\textbf{x}}+{\hat{a}}_{\textbf{k}}^{\dagger}{\cal R}_{k}^{*}\left(\eta\right)e^{-i\textbf{k}\cdot\textbf{x}}\right].
\end{equation}
In the above expression, $\hat{a}_{\textbf{k}}$  and ${\hat{a}}_{\textbf{k}}^{\dagger}$ are the usual creation and annihilation operators which satisfy the commutation relation $[\hat{a}_{\textbf{k}_1}, {\hat{a}}_{\textbf{k}_2}^{\dagger}]=\delta^{(3)}({\bf k}_1-{\bf k}_2)$ and annihilate the vacuum $|0\rangle$.  Plugging this in the equation of motion obtained from varying the quadratic action,
\begin{equation}
S = \frac{1}{2}\int{d\eta d^3x}a^2 \frac{\dot{\phi}^2}{H^2}\left[{\mathcal{R}'}^2-\left(\partial_i \mathcal{R}\right)^2\right],
\end{equation}
where prime denotes differentiation with respect to the conformal time $\eta$, and introducing the canonically normalized Mukhanov-Sasaki variable $v=z{\cal R} = a\delta\phi$, we find
\begin{equation}
v''_{k}+\left( k^{2}-\frac{z''}{z} \right)v_{k}=0,\label{Mukhanov-Sasaki}
\end{equation}
where $z=\frac{a\dot{\phi}}{H}$.  Note that the mode functions of the field perturbations $\delta\phi$ and $\delta\pi$ are given by the components of
\begin{equation}\label{mode-fxn}
\delta\Phi = \left( \frac{v}{a}, \frac{d}{dN} \left( \frac{v}{a} \right) \right).
\end{equation}

The dimensionless scalar power spectrum associated to the curvature perturbation is found from correlation of the Fourier modes $ \hat{\cal R}_{k}=\hat{a}_{\textbf{k}}{\cal R}_{k}+{\hat{a}}_{\textbf{k}}^{\dagger}{\cal R}_{k}^{*} $, evaluated at the end of inflation $\eta\to0$
\begin{equation}
\langle 0 | \hat{\mathcal{R}}_{{\bf k}_1} \hat{\mathcal{R}}_{{\bf k}_2} |0\rangle = \frac{\left(2\pi\right)^{2}}{2k^{3}} \mathcal{P}_{\cal R}(k) \delta^{(3)} \left(\textbf{k}_1+{\textbf{k}_2}\right),
\end{equation}
and is given by
\begin{equation}
\mathcal{P}_{\cal R}\left(k\right)=\frac{k^{3}}{2\pi^{2}}\left|{\cal R}_{k}\right|^{2}=\frac{k^{3}}{2\pi^{2}}\left|\frac{v_{k}}{z}\right|^{2}.\label{def:power spectrum}
\end{equation}
Thus the power spectrum is obtained by solving Eq.~\eqref{Mukhanov-Sasaki} for the mode function $v_{k}$.  In the quasi de~Sitter approximation that we use throughout, we have $a\left(\eta\right)=\frac{-1}{H\eta}$.  We now consider the pre- and post-transition cases $v^+$ and $v^-$ in turn.

In the slow-roll phase before the transition ($\phi>\phi_0$), we have $\frac{z^{\prime\prime}}{z}\simeq \frac{2}{\eta^{2}}$. So in this phase, the solution to the Mukhanov-Sasaki equation which corresponds to Bunch-Davis vacuum, $v_{k}={e^{-ik\eta}}/{\sqrt{2k}}$, in the remote past, is given by
\begin{equation}
v_{k}^{+}(\eta)=\frac{1}{\sqrt{2k}}\left(1-\frac{i}{k\eta}\right)e^{-ik\eta}.\label{vplus}
\end{equation}

The evolution of the Mukhanov-Sasaki variable after the slope change ($\phi<\phi_0$) is controlled by $\frac{z''}{z}$, as well. 
If we denote the conformal time coordinate 
at which the transition occurs by $\eta_{0}$, 
we have $\eta=\eta_0 e^{-n}$, where variations of $H$ are ignored due to the quasi de~Sitter approximation.  Then using Eqs.~\eqref{SR-background} and \eqref{USR-background} we find the function $z(\eta)$ to be
\begin{equation}
z = \frac{\beta D_0}{H\eta_0} \left[ \left(\frac{\eta}{\eta_0}\right)^2 - \frac{\eta_0}{\eta} \right] \Theta(\eta-\eta_0) - \frac{\pi_0}{H\eta},
\end{equation}
and that its derivative is given by
\begin{equation}
z' = \frac{\beta D_0}{H\eta_0^2} \left[ 2\frac{\eta}{\eta_0} +\left(\frac{\eta_0}{\eta}\right)^2 \right] \Theta(\eta-\eta_0) + \frac{\pi_0}{H\eta^2},
\end{equation}
Hence we have
\begin{equation}
z'' = \frac{\beta D_0}{H\eta_0^2} \left[ 2\frac{\eta}{\eta_0} +\left(\frac{\eta_0}{\eta}\right)^2 \right] \delta(\eta-\eta_0) + \text{``finite terms''}.
\end{equation}
Integrating Eq.~\eqref{Mukhanov-Sasaki} around $\eta_0$ we find the discontinuity in $v'$ across the transition to be
\begin{equation}\label{disc-v}
v_k'^-(\eta_0) - v_k'^+(\eta_0) = \frac{3f}{\eta_0} v^-_k(\eta_0),
\end{equation}
where 
\begin{equation}
f = -\frac{\beta D_0}{\pi_0} = \frac{\alpha-\beta}{\alpha}
\end{equation}
is the fractional change in the slope of the potential.  The Bunch-Davies initial condition can only be applied to $v^+$, while $v^-$ has to be found by matching the general solution
\begin{equation}
v_{k}^{-}\left(\eta\right)=\frac{1}{\sqrt{2k}}\left[\alpha_{k}\left(1-\frac{i}{k\eta}\right)e^{-ik\eta}+\beta_{k}\left(1+\frac{i}{k\eta}\right)e^{ik\eta}\right]\label{vminus}
\end{equation}
to $v^+$ across the discontinuity using Eq.~\eqref{disc-v}, which gives the Bogoliubov coefficients
\begin{equation}
\alpha_{k}=1 - \frac{3if}{2}\frac{k_{0}}{k}\left(1+\frac{k_{0}^{2}}{k^{2}}\right),\label{alpha-k}
\end{equation}
\begin{equation}
\beta_{k}=\frac{3if}{2}\frac{k_{0}}{k}\left(1+\frac{ik_{0}}{k}\right)^{2} e^{2i\frac{k}{k_0}}.\label{beta-k}
\end{equation}
Here $k_{0}$ refers to the mode which exits the Hubble radius at the transition time. 

Substituting the values $\alpha_{k}$ and $\beta_{k}$ in Eq.~\eqref{vminus}, one finds the power spectrum at the end of inflation $k\eta\rightarrow 0$ from Eq.~\eqref{def:power spectrum} to be~\cite{Starobinsky:1992ts}
\begin{equation}\label{power spectrum}
\begin{aligned}
\mathcal{P}_{\cal R}\left(k\right) &= \left(\frac{H}{2\pi\beta}\right)^2 \left|\alpha_{k}-\beta_{k}\right|^2 \\
&= \left(\frac{H}{2\pi\beta}\right)^{2} \left\{ 1 + \frac{9}{2}f^{2} \frac{k_0^2}{k^{2}} \left(1+\frac{k_0^2}{k^{2}}\right) \left[ \left(1+\frac{k_0^2}{k^{2}}\right) + \left(1-\frac{k_0^2}{k^{2}}\right) \cos \frac{2k}{k_0} - \frac{2k_0}{k}\sin \frac{2k}{k_0} \right] \right. \\
& \left. + 3f \frac{k_0}{k} \left[ \left(1-\frac{k_0^2}{k^{2}}\right)\sin \frac{2k}{k_0} + \frac{2k_0}{k} \cos \frac{2k}{k_0} \right] \right\}.
\end{aligned}
\end{equation}
This spectrum depends on the wavenumber through $k/k_0$ and we have
\begin{equation}\label{standard-power-limits}
\mathcal{P}_{\cal R}(k) = 
\begin{cases}
\left( \dfrac{H}{2\pi\alpha} \right)^2, & k\ll k_0, \\
\left( \dfrac{H}{2\pi\beta} \right)^2, & k\gg k_0,
\end{cases}
\end{equation}
which exhibits an amplification of power on short scales by a factor of $(\alpha/\beta)^2$.  This enhancement occurs around $k=k_0$ and is followed by a series of oscillations.  In fact, the enhancement itself is a natural consequence of the fact that in the early and late slow-roll phases the power is inversely proportional to the slow-roll parameter $\varepsilon_1$.
This behavior 
has been noticed in 
previous works 
that analyze the power spectrum numerically or analytically \cite{Starobinsky:1992ts}.

The mode functions obtained in this section will be of use in determining the stochastic noise.  Before embarking on that calculation, we review the stochastic formalism in the next section.

\section{Review of the Stochastic Formalism}\label{sec:review-stochastic}

The stochastic approach is based on smoothing the fields on physical scales smaller than $1/\sigma H$, where $\sigma\ll1$ is some cutoff parameter, typically of order $0.1$ or $0.01$.  The coarse-grained fields $\phi$ and $\pi$ satisfy the Langevin equation~\cite{Grain:2017dqa}
\begin{align}
\frac{d\phi}{dN} &= \pi + \xi_\phi, \label{Langevin-phi} \\
\frac{d\pi}{dN} &= -(3-\varepsilon_1)\pi - \frac{V_{,\phi}}{H^2} + \xi_\pi. \label{Langevin-pi}
\end{align}
We can write these equations in the compact form $d\Phi_i/dN = D_i + \xi_i$ for the fields $\Phi_i \in \{ \phi, \pi \}$ and call $D_i$ the drift coefficients.  The correlation of the Gaussian noises $\xi_i$ is given by \cite{Grain:2017dqa}
\begin{equation}
\langle \xi_i(N) \xi_j(N') \rangle = \Xi_{ij}(N) \delta(N-N'),
\end{equation}
where
\begin{equation}\label{noises}
\begin{aligned}
\Xi_{ij}\left(N\right) &= \frac{d\ln k_{\sigma}}{dN} \frac{k^{3}_{\sigma}}{2\pi^{2}} \delta\Phi_i \left(k_\sigma, N\right) \delta\Phi^{*}_{j}\left(k_\sigma, N\right) \\
&= (1-\varepsilon_1) \frac{k^{3}_{\sigma}}{2\pi^{2}} \delta\Phi_i \left(k_\sigma, N\right) \delta\Phi^{*}_{j}\left(k_\sigma, N\right),
\end{aligned}
\end{equation}
in which $\delta\Phi_i$ is the mode function corresponding to the perturbations of the field $\Phi_i$ defined in Eq.~\eqref{mode-fxn}, and $k_{\sigma}=\sigma aH$ is the comoving cutoff scale.  Since we are in the quasi de~Sitter regime, we drop the $\varepsilon_1$ term in the forthcoming calculation of $\Xi$.

In the stochastic $\delta N$ formalism, we consider a region $\Omega$ in the phase space whose boundary $\partial\Omega$ is comprised of two components.  One component, $\partial\Omega_e$, determines the surface of the end of inflation in the phase space.  The other component, $\partial\Omega_{\rm UV}$, acts like a cutoff and is a reflecting boundary that is required to prohibit the inflaton from reaching UV regions.  The probability $P(\Phi,N)$ of finding the field value $\Phi$ at the time $N$ satisfies the Fokker-Planck equation
\begin{equation}
\frac{\partial P}{\partial N} = {\cal L} P = \frac{\partial J_i}{\partial\Phi_i} = \frac{\partial}{\partial\Phi_i} \left[ D_i P + \frac12 \frac{\partial}{\partial\Phi_j} (\Xi_{ij} P) \right],
\end{equation}
where $\cal L$ is the Fokker-Planck operator and $J$ is the probability current.  The appropriate boundary conditions are of absorbing type on $\partial\Omega_e$ and of reflecting type on $\partial\Omega_{\rm UV}$ \cite{Gardiner}:
\begin{equation}
P \big|_{\partial\Omega_e} = 0, \qquad n_i J_i \big|_{\partial\Omega_{\rm UV}} = 0, \qquad \forall N>0.
\end{equation}
where $n$ is the normal to the boundary.  Of course, one also needs an initial condition (in time) to specify the initial distribution $P(\Phi, N=0)$ at $N=0$.

A central quantity is the first boundary crossing ${\N}(\Phi)$, which measures the number of $e$-folds in each realization between an initial field value $\Phi$ and the first time $\partial\Omega$ is crossed.  The probability distribution associated with ${\N}(\Phi)$, denoted by $P_{\N}(\Phi, N)$ or $P_{\Phi}(N)$, satisfies the adjoint Fokker-Planck equation:
\begin{equation}
\frac{\partial P_\N}{\partial N} = {\cal L}^\dag P_\N = \left[ D_i \frac{\partial}{\partial\Phi_i} + \frac12 \Xi_{ij} \frac{\partial^2}{\partial\Phi_i \partial\Phi_j} \right] P_\N,
\end{equation}
with absorbing and reflecting boundary conditions on $\partial\Omega_e$ and $\partial\Omega_{\rm UV}$ respectively given by~\cite{Gardiner}:
\begin{equation}\label{P_N b.c.}
P_\N \big|_{\partial\Omega_e} = 0, \qquad n_i \Xi_{ij} \partial_j P_\N \big|_{\partial\Omega_{\rm UV}} = 0, \qquad \forall N>0.
\end{equation}
Since the trajectory is assumed to start from $\Phi$, the initial condition on $P_\N$ is
\begin{equation}
P_\N(\Phi, N=0) = \delta(\Phi\in\partial\Omega_e),
\end{equation}
where the delta function is meant to blow up for $\Phi\in\partial\Omega_e$ and is zero at all interior points.

The Fourier transform of $P_{\N}$,
\begin{equation}
\chi(\Phi, t) = \int_0^\infty P_\N(\Phi, N) e^{itN} dN,
\end{equation}
also known as the characteristic function, is an eigenfunction of the adjoint Fokker-Planck operator:
\begin{equation}
{\cal L}^\dag \chi = -it\chi.
\end{equation}
Its boundary conditions are inherited from $P_\N$~\cite{Gardiner}, namely,
\begin{equation}\label{chi b.c.}
\chi \big|_{\partial\Omega_e} = 1, \qquad n_i \Xi_{ij} \partial_j \chi \big|_{\partial\Omega_{\rm UV}} = 0, \qquad \forall t.
\end{equation}
$\chi$ does not require an initial condition, but the normalization of $P_\N$ implies that $\chi(\Phi, 0)=1$ for all $\Phi$.

The moments of $\cal N$ defined by
\begin{equation}
\langle {\cal N}^k \rangle = \int_0^\infty P_\N(\Phi, N) N^k dN 
\end{equation}
are of interest; they provide the statistics of first passage time starting from the point $\Phi$ in the phase space.  The characteristic function can be used to compute these moments as follows:
\begin{equation}
\chi(\Phi, t) = \sum_{k=0}^\infty \int_0^\infty P_\N(\Phi, N) \frac{(itN)^k}{k!} dN = \sum_{k=0}^\infty \langle {\cal N}^k \rangle \frac{(it)^k}{k!},
\end{equation}
yielding
\begin{equation}\label{moments}
\langle {\cal N}^k \rangle = \frac{1}{i^k} \left. \frac{d^k}{dt^k} \chi(\Phi,t) \right|_{t=0}.
\end{equation}

To investigate the large-$N$ behavior of the PDF $P_\N$, we look at the poles of its Fourier transform $\chi$.  Note that since $P_\N(N)$ is zero for negative $N$ then, according to the Titchmarsh theorem, all poles of $\chi(t)$ lie in the lower half of the complex $t$-plane. 
The contribution of an exponentially falling tail $P_\N(N) \sim pe^{-\Lambda N}$ to $\chi$ consists of a simple pole that diverges as $it\to\Lambda$:
\begin{equation}
\chi(t) = \int_0^\infty P_\N e^{itN} dN = \frac{p}{it-\Lambda} + \text{``finite as $it\to\Lambda$''}.
\end{equation}
Therefore, the smallest pole (in $it$) of $\chi$, which we denote by $\Lambda_0$, corresponds to the asymptotic exponential decay rate of $P_\N(N)$.  An alternative approach to study the tail of the PDF, by direct investigation of the $P_{\cal N}$ is outlined in Appendix~\ref{app:tail}.

\section{The Stochastic Noises}\label{sec:noise}

We now move on to compute the noise correlation matrix. 
Using Eqs.~\eqref{mode-fxn} and \eqref{vplus}, we obtain the mode functions of field perturbations:
\begin{equation}
\delta\phi_k^+(\eta) = \frac{-H\eta}{\sqrt{2k}} \left(1-\frac{i}{k\eta}\right) e^{-ik\eta}, \qquad
\delta\pi_k^+(\eta) = \frac{-ikH\eta^2}{\sqrt{2k}} e^{-ik\eta}.
\end{equation}
Then using Eq.~\eqref{noises} we find
\begin{equation}
\Xi = \left( \frac{H}{2\pi} \right)^2
\begin{pmatrix}
1+\sigma^2 				& -\sigma^2(1-i\sigma)\\
-\sigma^2(1+i\sigma)	& \sigma^4
\end{pmatrix}.
\end{equation}
From now on, we will drop the subleading terms in $\sigma$ which also eliminates the imaginary part appearing in the off-diagonal elements of $\Xi$ (at any rate the imaginary part would not correspond to any classical noise~\citep{Grain:2017dqa}) and write:
%
\begin{equation}\label{noise-I}
\Xi_{\rm I} = \left(\frac{H}{2\pi}\right)^{2} \begin{pmatrix}
1 & -\sigma^2 \\
-\sigma^2 & \sigma^4
\end{pmatrix},
\end{equation}
where the subscript I means ``phase I'' by which we refer the pre-transition slow-roll stage.  The noise matrix has a zero eigenvalue which implies that the noises in the $\phi$ and $\pi$ directions are totally anti-correlated. This property of the noises in the slow-roll regime is well known~\cite{Grain:2017dqa}.

We now calculate the noise for the post-transition phase.  Expanding the exponential in $\alpha_{k_{\sigma}}$ and $\beta_{k_{\sigma}}$ in powers of $\sigma$, we find
\begin{equation}\label{delta-phi-}
\delta\phi_{k_{\sigma}}^{-}\left(\eta\right)=\frac{-H\eta}{\sqrt{2k_{\sigma}}}\frac{i}{\sigma}\left[\left(1+\frac{\sigma^{2}}{2}+O(\sigma^4)\right)\left(\alpha_{k_\sigma}-\beta_{k_\sigma}\right)+i\left(\frac{\sigma^{3}}{3}+O(\sigma^4)\right)\left(\alpha_{k_\sigma}+\beta_{k_\sigma}\right)\right]
\end{equation}
and
\begin{equation}\label{delta-pi-}
\delta\pi_{k_{\sigma}}^{-}(\eta) = \frac{-iH\eta^{2}k_{\sigma}}{\sqrt{2k_{\sigma}}}\left[\left(1-\frac{\sigma^{2}}{2}+O(\sigma^4)\right)\left(\alpha_{k_{\sigma}}-\beta_{k_{\sigma}}\right)+i\sigma\left(1-\frac{\sigma^{2}}{6}+O(\sigma^4)\right)\left(\alpha_{k_\sigma}+\beta_{k_\sigma}\right)\right].
\end{equation}
As stated in Section~\ref{perturbations}, the power spectrum after the transition is controlled by the ratio $k/k_0$, so let us define a new variable 
\begin{equation}\label{def:lambda}
\lambda=\frac{k_{\sigma}}{k_{0}}=\sigma e^{n},
\end{equation}
based on which different phases will be identified.  The above coefficients $\alpha_{k_{\sigma}}$ and $\beta_{k_{\sigma}}$ can be expressed in terms of $\lambda$ as 
\begin{equation}\label{alpha-beta-lambda}
\alpha_{k_{\sigma}}\pm\beta_{k_{\sigma}}=1-\frac{3if}{2}\frac{1}{\lambda}\left[1+\frac{1}{\lambda^2}\mp\left(1+\frac{i}{\lambda}\right)^2 e^{2i\lambda}\right].
\end{equation}
We next investigate the two opposite limits $\lambda\ll1$ and $\lambda\gtrsim1$ in turn.

Let us consider the post-transition period when $\lambda\ll1$, which starts just after the slope change when $\lambda=\sigma\ll1$ and ends well before $\lambda\sim1$, i.e., well before $n\sim-\log\sigma$ (for $\sigma=0.01$ this phase lasts for about 2--3 $e$-folds).  In this phase
\begin{equation}
\alpha_{k_{\sigma}}-\beta_{k_{\sigma}}=1-f+O\left(\lambda^{2}\right), \qquad 
\alpha_{k_{\sigma}}+\beta_{k_{\sigma}} = \frac{-3if}{\lambda^{3}} \left[ 1+O(\lambda^{2})\right].
\end{equation}
The noise amplitudes are given by
\begin{equation}\label{noise-II&III}
\begin{split}
&\Xi=\left(\frac{H}{2\pi}\right)^2 \times \\
&\begin{pmatrix}
\left[1-f\left(1-e^{-3n}\right)\right]^2 & \left[1-f\left(1-e^{-3n}\right)\right] \left[ -3f e^{-3n} + (1-f)\sigma^2 \right] \\
\left[1-f\left(1-e^{-3n}\right)\right] \left[ -3f e^{-3n} + (1-f)\sigma^2 \right] & \left[ -3f e^{-3n} + (1-f)\sigma^2 \right]^2
\end{pmatrix}.
\end{split}
\end{equation}
This expression is valid for the entire period $\lambda\ll1$, but it is more convenient to split this period to two parts: ``phase II'' where the first term in $\Xi_{\pi\pi}$ dominates, i.e.,
\begin{equation}\label{def:phaseII}
f e^{-3n} \gg (1-f)\sigma^2 \quad \Longleftrightarrow \quad \lambda^3 \ll \frac{f\sigma}{1-f},
\end{equation}
and ``phase III'' where the second term dominates, i.e.,
\begin{equation}\label{def:phaseIII}
f e^{-3n} \ll (1-f)\sigma^2 \quad \Longleftrightarrow \quad \lambda^3 \gg \frac{f\sigma}{1-f}.
\end{equation}
We note that when $\beta\to0$ (i.e., $f\to1$) Eq.~\eqref{def:phaseIII} becomes inconsistent with $\lambda\ll1$ and hence we won't have a phase III.  In other words, phase III exists only if $\sigma\ll\beta/(\alpha-\beta)$.  We find that for phase II the noise amplitudes~\eqref{noise-II&III} reduce to
\begin{equation}\label{noise-IIA}
\begin{aligned}
\Xi_{\rm II} &= \left(\frac{H}{2\pi}\right)^2 \begin{pmatrix}
\left[1-f\left(1-e^{-3n}\right)\right]^2 & -3f\left[1-f\left(1-e^{-3n}\right)\right]e^{-3n}\\
-3f\left[1-f\left(1-e^{-3n}\right)\right] e^{-3n}& 9f^2 e^{-6n}
\end{pmatrix} \\
&= \left(\frac{H}{2\pi\pi_0}\right)^2 \begin{pmatrix}
\langle\pi\rangle^2 & 3\beta\langle\pi\rangle \langle D\rangle \\
3\beta\langle\pi\rangle \langle D\rangle & 9\beta^2 \langle D\rangle^2
\end{pmatrix}.
\end{aligned}
\end{equation}
where in the second line we have rewritten it in terms of the background value $\langle\pi\rangle$ given by Eq.~\eqref{USR-background}.
Similarly, for phase III we have
\begin{equation}\label{noise-IIIA}
\Xi_{\rm III}=\left(\frac{H}{2\pi}\right)^2 \begin{pmatrix}
\left[1-f\left(1-e^{-3n}\right)\right]^2 & \left[1-f\left(1-e^{-3n}\right)\right] (1-f)\sigma^2 \\
\left[1-f\left(1-e^{-3n}\right)\right] (1-f)\sigma^2 & (1-f)^2\sigma^4
\end{pmatrix}.
\end{equation}

\begin{figure}
\centering
\includegraphics[scale=1]{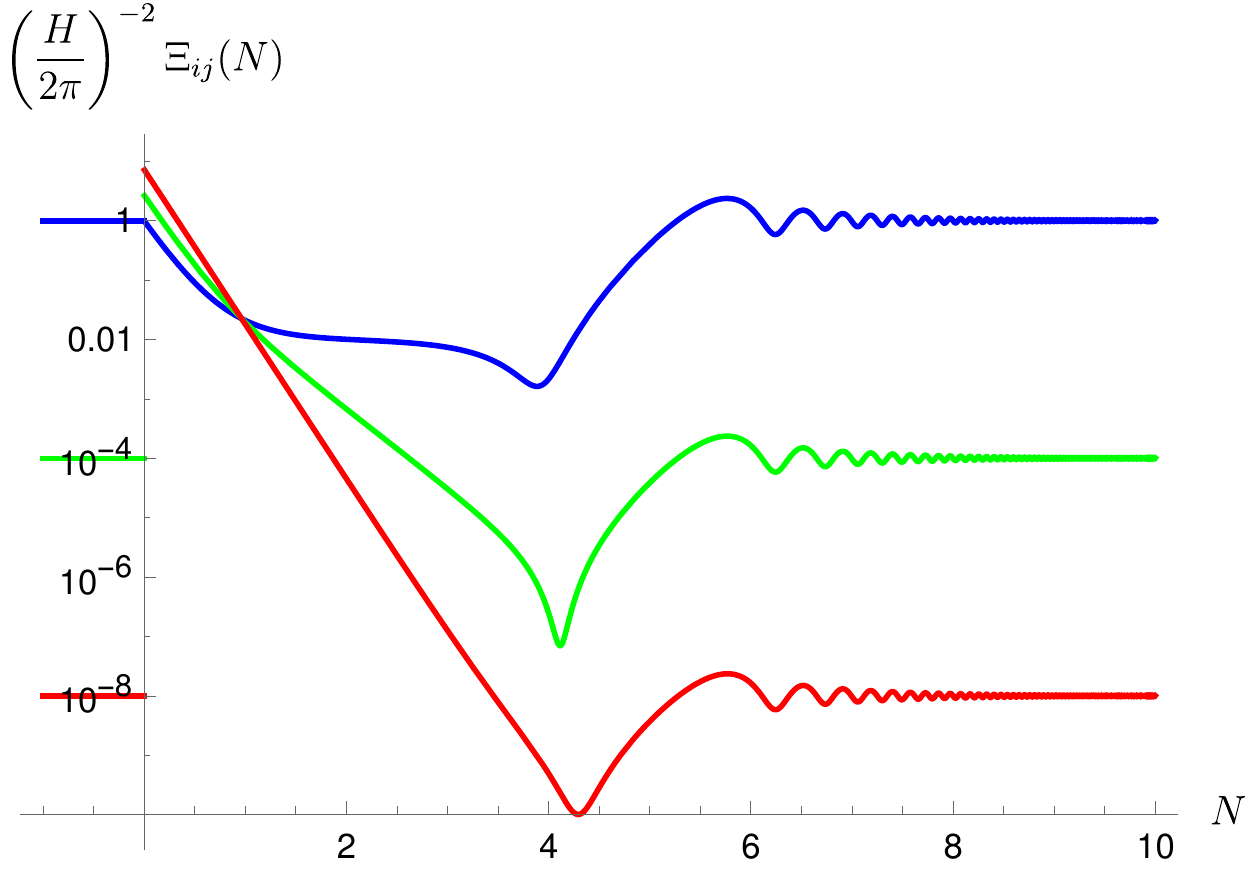}
\caption{Noise correlation matrix elements $\Xi_{ij}$ in units of $(\frac{H}{2\pi})^2$ as functions of $N$: $\Xi_{\phi\phi}$ (blue), $-\Xi_{\phi\pi}$ (green) and $\Xi_{\pi\pi}$ (red). Parameter values are $\alpha=0.1$, $\beta=0.01$ and $\sigma=0.01$.  There is no plateau in $\Xi_{\pi\pi}$ after the exponential fall-off, indicating the absence of phase III.  This is because, according to Eqs.~\eqref{def:phaseII} and \eqref{def:phaseIII}, the transition from phase II to III would have occurred at $n\approx4.2$ corresponding to $\lambda=0.65$ which breaks the assumption $\lambda\ll1$.
}\label{fig:noise}
\end{figure}

As we mentioned above, phase III does not occur in certain circumstances, including the parameters we are interested in.  Therefore, we will not elaborate on it further, except for making the remark that unlike the other phases where the noises in the $\phi$ and $\pi$ directions are anti-correlated, these noises are correlated in phase III. 
Regarding the noise in phase II, the situation is different from the slow-roll regime, where $\xi_\pi$ can be ignored.  The main result of this analysis is the following: The ratio of the noise term $\xi_i$ in the Langevin equations~\eqref{Langevin-phi} and \eqref{Langevin-pi} to the drift term $D_i$ is controlled by
\begin{equation}
\dfrac{\sqrt{\langle \xi_\phi^2 \rangle}}{\langle D_\phi \rangle} = \dfrac{\dfrac{H}{2\pi} \dfrac{\langle \pi \rangle}{\pi_0} \delta(0)}{\langle \pi \rangle},
\qquad \text{and} \qquad
\dfrac{\sqrt{\langle \xi_\pi^2 \rangle}}{\langle D_\pi \rangle} = \dfrac{\dfrac{H}{2\pi} \dfrac{3\beta \langle D \rangle}{\pi_0} \delta(0)}{ 3\beta \langle D \rangle},
\end{equation}
which happen to be identical now.
This implies that the $\pi$ noise is as important as the $\phi$ noise, and it can be neglected only when the field is far from the transition. The jump in the $\pi$ noise across the transition is inherited from the discontinuity in $v'$. We expand on this view in Appendix~\ref{app:transition-noises} and demonstrate that the velocity noise becomes suddenly more pronounced right after every transition with a sudden change in $\pi$. Therefore, one has to consider also the spread of distribution of velocities in the presence of a transition to a non-slow-roll regime. This is the most intriguing part of this result.  The noise amplitude in phase II may lead to some changes in the power spectrum relative to the expression given in Eq.~\eqref{power spectrum}.  We will look into this case in Sections~\ref{sec:perturbative} and \ref{sec:nonperturbative}.
As soon as $n\sim-\log\sigma$, i.e., $\lambda\sim1$, the above approximation breaks down and we enter phase IV, defined by $\lambda\gtrsim1$.  In this limit we find from Eq.~\eqref{alpha-beta-lambda} that $\alpha_{k_{\sigma}}\approx1$ and $\beta_{k_{\sigma}}\approx0$, so that the terms involving $\alpha_{k_{\sigma}}+ \beta_{k_{\sigma}}$ in Eqs.~\eqref{delta-phi-} and \eqref{delta-pi-} become negligible; thus
\begin{equation}\label{noise-IV}
\Xi_{\rm IV} = |\alpha_{k_{\sigma}}-\beta_{k_{\sigma}}|^{2} \left(\frac{H}{2\pi}\right)^{2}
\begin{pmatrix}
1 & -\sigma^2 \\
-\sigma^2 & \sigma^4
\end{pmatrix}
= \beta^2 {\cal P}_{\cal R}(k_\sigma)
\begin{pmatrix}
1 & -\sigma^2 \\
-\sigma^2 & \sigma^4
\end{pmatrix}.
\end{equation}
In the second slow-roll regime when $\lambda\gg1$, the noise amplitude is similar to that of the earlier slow-roll regime~\eqref{noise-I}, as one would expect. The elements of noise correlations, $\Xi_{ij}\left(n\right)$, corresponding to $f=0.9$, are illustrated in Figure~\ref{fig:noise}. It is easy to see that the velocity noise is as crucial as the field noise in about two $e$-folds after the transition. Additionally, the noise hierarchy is at odds with the two limiting cases. It is conceivable that the noise effects are captured in a stochastic analysis. 

\section{Statistics of Cosmological Fields After the Transition: Perturbative Analysis}\label{sec:perturbative}

The stochastic formalism is applicable to modes $k$ that are classic during the era under study, i.e., $k\ll k_\sigma$.  In particular, to apply the $\delta N$ formalism, we need this condition to hold all the way from the initial to the final hypersurface.  Our goal is to study the effect of the noises of the transition era, so we require $k\ll k_\sigma(N)$ for $N$ in this era.  Given that $\lambda\ll1$ in the immediate aftermath of the transition (namely, phase II), it follows from Eq.~\eqref{def:lambda} that $k\ll k_\sigma\ll k_0$.  This implies that  $k$ has exited horizon much before the transition.  In terms of the initial hypersurface $\phi$, from which $\cal N$ is counted, this means that we have to evaluate $\chi$ 
at a $\phi$ that is far enough to the right of $\phi_0$.  To simplify the calculations, we restrict our attention to phase II only, putting the right boundary at the transition point $\phi_0$, and set the starting point $\phi$ right on this boundary.  This means that we set aside the effect of the noises of all the other phases, which are either slow-roll attractor stages or have tiny velocity noises.  Those effects are already studied in the literature, so we concentrate on phase II with its novel form of noise.  In the sequel, we carry our calculations with a generic starting point $\phi$, but eventually set $\phi=\phi_0$.

As noted above, in the rest of this paper we concentrate on phase II, where the noise is given by Eq.~\eqref{noise-IIA}.  This is a time-dependent noise $\Xi(n)$ and is calculated based on a particular background trajectory, namely, the attractor trajectory highlighted in Figure~\ref{fig:phase} (for example, the mode function $v$ in Eq.~\eqref{Mukhanov-Sasaki} depends on background quantities such as $\dot\phi$ through $z$).  This expression for noise is therefore most reliable for realizations that do not wander too far from the background trajectory.  In such cases, we may replace $n$ in $\Xi(n)$ by $\pi$ using the background equation of motion~\eqref{USR-background}.\footnote{In the absence of quantum diffusion, the analysis of a time dependent function is simple since it is directly related to a point in phase space and evolves along a reference classical trajectory. In the stochastic picture, however, this one-to-one correspondence between time variable and phase space variables is lost, which is what makes solving the Fokker-Planck equation with time dependent noises challenging. On the other hand the noises are obtained from Mukhanov-Sasaki equation. We could have decided to rewrite this equation in terms of the background $\phi$ or $\pi$ as the independent variable, from that $\Xi\left(\phi\right)$ or $\Xi\left(\pi\right)$ would have been found.}  This turns $\Xi\left(n\right)$ into $\Xi\left(\pi\right)$,
\begin{equation}\label{noise-IIB}
\Xi = \left( \frac{H}{2\pi\pi_0} \right)^2
\begin{pmatrix}
\pi^2 & 3\beta\pi D(\pi) \\
3\beta\pi D(\pi) & 9\beta^2 D(\pi)^2
\end{pmatrix},
\end{equation}
which is the noise we use in the sequel.  Comparing with the second line of Eq.~\eqref{noise-IIA}, this corresponds to removing the expectation values, which clearly leads to a different expression for the noises away from the background trajectory.  We would like to stress that neither of the two expressions \eqref{noise-IIA} and \eqref{noise-IIB} can be claimed as the correct noise of stochastic inflation, so our choice of~\eqref{noise-IIB} is a matter of choice.  We could have equally chosen to express $\Xi$ in terms of the background $\phi$ and that would yield yet another different expression.  The virtue of working with a noise that is a function of the phase space variables ($\phi$ or $\pi$ or both, as opposed to $n$) is the subsequent simplicity in dealing with time-independent Fokker-Planck equations, as we see later.  The reason we have chosen $\pi$ over $\phi$ is also the simpler algebraic form of the noise.  Let us further emphasize that for small diffusion, the difference between any of these expressions is negligible.  Indeed, the stability analysis performed in~\cite{Pattison:2018bct} shows that the USR regime in the phase space of the Starobinsky model is unstable and lasts for only a few number of $e$-folds.  It is therefore highly unlikely that a typical trajectory (close to the background phase space path) has so high fluctuations to take it far from the average trajectory in such a short span of time.

As a measure of the stochastic effects we define another dimensionless parameter $\kappa$ as
\begin{equation}
\kappa = \frac12 \left( \frac{H}{2\pi\pi_0} \right)^2 = \frac12 \left( \frac{H^2}{2\pi\dot\phi_0} \right)^2.
\end{equation}
This would be useful when working on diffusion effects.\footnote{Note the difference between the definition of $\kappa$ here and Ref.~\cite{Firouzjahi:2018vet} which defines $\kappa = 3e^{3N_c}H^2/2\pi\dot\phi_0 = 3H^2/2\pi\dot\phi_e = 3\sqrt{{\cal P}_e}$.} This variable determines whether inflation is in the drift- or in the diffusion-dominated regime.\footnote{The inverse of $\kappa$ quantifies the ratio of the rate of drift to the rate of diffusion in $\phi$ evolution. In the context of transport phenomena in a continuum, it corresponds to the {\it P\'eclet} number. It has also applications beyond transport phenomena, as a general measure for the relative importance of the random fluctuations and of the systematic average behavior in mesoscopic systems}  Let us have an estimation for the possible values of $\kappa$ in different regions of inflationary potential. The current bounds on inflationary Hubble parameter imposes $H\lesssim 10^{-5}M_{\rm Pl}$. On the other hand, in order to have the initial slow-roll in Starobinsky model, $\alpha$ should be taken small enough, $\alpha\ll M_{\rm Pl}$. Therefore, $\kappa\ll 1$ as long as $\alpha\gtrsim 10^{-5}M_{\rm Pl}$.  When the slow-roll conditions are satisfied, $2\kappa\ll1$ is a good approximation for the power spectrum on large scales.  The condition $\kappa\gg1$ shows that the quantum jumps are significant and control the evolution of the inflation field.

The adjoint Fokker-Planck equation reads
\begin{equation}
\kappa \left[ \pi^2 \frac{\partial^2\chi}{\partial\phi^2} + 6\beta D\pi \frac{\partial^2\chi}{\partial\phi \partial\pi} + 9\beta^2D^2 \frac{\partial^2\chi}{\partial\pi^2} \right] + \pi \frac{\partial\chi}{\partial\phi} + 3\beta D \frac{\partial\chi}{\partial\pi} + it\chi = 0.\label{adjoint Fokker-Planck equation}
\end{equation}
In a drift-dominated regime, $\kappa$ defines a small parameter, useful for our arguments based on a perturbative expansion. 
In what comes next we decompose $\mathcal{L}^{\dagger}$ into drift and diffusion parts as
\begin{equation}
\mathcal{L}^{\dagger}=\mathcal{L}^\dag_{(0)}+\mathcal{L}^\dag_{(1)},\ \ \ \ \mathcal{L}^\dag_{(0)}\equiv\pi \frac{\partial}{\partial\phi}+3\beta D\frac{\partial}{\partial\pi},\ \ \ \ \  \mathcal{L}^\dag_{(1)}\equiv\frac{1}{2}\Xi_{ij}\frac{\partial^2}{\partial\Phi_i\partial\Phi_j}.
\label{L decomposition}
\end{equation}
We further assume a perturbative expansion for $\chi$ in the form of $\chi=\chi_{\left(0\right)}+\chi_{\left(1\right)}+\chi_{\left(2\right)}+\ldots$, for the small $\kappa$ limit. The equations governing the $i$-th term in the expansion are obtained by substituting this form into Eq.~\eqref{adjoint Fokker-Planck equation}. Let us denote the absorbing boundary $\partial\Omega_e$ in phase space by $\phi=\phie$, which corresponds to the end of transition effects and hence effectively indicates the end of inflation.  Then the leading order term, $\chi_{(0)}$, would be the solution of the following partial differential equation (PDE),
\begin{equation}
\left[\pi\frac{\partial}{\partial\phi}+3\beta D\frac{\partial}{\partial\pi}\right]\chi_{\left(0\right)}=-it\chi_{\left(0\right)},\ \ \ \ \chi_{\left(0\right)}\left(\phie,\pi;t\right)=1\label{chi0}.
\end{equation}
We also find that the $i$-th term in the expansion satisfies the following set of iterative equations
\begin{equation}
\left( \mathcal{L}_{(0)} + it \right) \chi_{\left(i\right)}=-\mathcal{L}_{(1)}\chi_{\left(i-1\right)},\qquad \forall i\geq1\label{chi(i neq 0)},
\end{equation}
and the boundary condition $\chi_{\left(i\neq 0\right)}\left(\phie,\pi;t\right)=0$.
 
We use the method of characteristics to find the solution of the linear first order PDE governing the leading order. A detailed derivation of this solution is given in Appendix~\ref{app:characteristics}; we summarize the results here. The solution $\chi_{(0)}\left(\phi,\pi\right)$ of Eq.~\eqref{chi0} is given by
\begin{equation}
\chi_{(0)}(\phi,\pi;t) = \left[\frac{D(\pi)}{W\left(Z\right)}\right]^{\frac{it}{3}}.\label{chi-0-solution}
\end{equation}
Here $W$ stands for the Lambert $W$ function (defined by $W(x)e^{W(x)}=x$) and\footnote{In Sections~\ref{sec:background} and \ref{sec:noise} we used $\eta$ to denote conformal time.  We no longer do so.}
\begin{equation}\label{def:Z-eta}
Z\equiv\frac{1}{e\beta}e^{-\frac{3}{\beta}\left(\eta-\phie\right)}, \qquad \text{where} \qquad
\eta\equiv\phi + \frac{\pi}{3}-\frac{\beta}{3}\ln \left(\beta D\right),
\end{equation}
is a function that is constant along the classical deterministic trajectories of Figure~\ref{fig:phase}.  When evaluated on the boundary curve $\phi=\phie$, $W(Z)$ reduces to the previously defined function $D(\pi)$, so that the boundary condition of Eq.~\eqref{chi0} is satisfied.  Note that $\eta$ is the first integral of motion in the absence of quantum diffusion. It, however, changes from one trajectory to another in Figure~\ref{fig:phase}; so it may be considered as a label for different classical paths.

For a flat potential, $\beta=0$, the PDE reduces to
 \begin{equation}
\left[\pi\frac{\partial}{\partial\phi}+3\pi\frac{\partial}{\partial\pi}\right]\chi_{\left(0\right)}=-it\chi_{\left(0\right)},\qquad \chi_{\left(0\right)}\left(\phie,\pi;t\right)=1\label{chi(0)},
\end{equation} 
and the solution is simply given by 
\begin{equation}
  \chi_{(0)}=\left(1+\frac{3(\phi-\phie)}{\pi}\right)^{\frac{-it}{3}} = \left(\frac{\eta-\phie}{\eta-\phi}\right)^{\frac{-it}{3}}
\label{chi0-beta0}.
\end{equation}
At the leading order, the mean number of $e$-folds, $\left\langle \mathcal{N}\right\rangle_{\left(0\right)}$, is given by $\frac{1}{3}\ln\left({D}/{W\left(Z\right)}\right)$ for arbitrary $\beta$; and by
\begin{equation}
\left\langle \mathcal{N}\right\rangle_{\left(0\right)} = \frac{1}{3}\ln \left( 1+\frac{3(\phi-\phie)}{\pi} \right) =\frac13 \ln \left(\frac{\eta-\phie}{\eta-\phi}\right)
\end{equation}
for the $\beta=0$ case. The PDF of the number of $e$-folds $\mathcal{N}$, and hence of coarse-grained $\mathcal{R}$, can be obtained by Fourier transforming the classical results~\eqref{chi-0-solution} and~\eqref{chi0-beta0},
\begin{equation}\label{PDF-delta}
P_{\mathcal{N}}^{(0)}\left(\phi, \pi; N \right)=\delta\left(N-\left\langle \mathcal{N}\right\rangle_{\left(0\right)}\right).
\end{equation}
As noted above, for the modes of interest the relevant result is obtained by using $\Phi_0$ in place of $\Phi$ in Eq.~\eqref{PDF-delta} and expressions given for $\left\langle \mathcal{N}\right\rangle_{\left(0\right)}$.  This includes replacing $D$ by $D_0$ and $W$ by $W_0$, where $W_0=W(Z_0)$ and $Z_0$ is obtained from evaluating Eq.~\eqref{def:Z-eta} at $\phi=\phi_0$ and $\pi=-\alpha$.

Let us now concentrate on the next order terms in the expansion.  One can put Eq.~\eqref{adjoint Fokker-Planck equation} into its canonical form by using $\eta$ defined in Eq.~\eqref{def:Z-eta} as a new variable, so that our variables are now $(\phi,\eta)$ instead of $(\phi,\pi)$.  The inverse transformation is given by
\begin{equation} 
\pi=-\beta\left[1+D(\phi,\eta)\right], \qquad D(\phi,\eta)=W\left(\frac{1}{e\beta}e^{-\frac{3}{\beta}\left(\eta-\phi\right)}\right).\label{transformation}
\end{equation}
Note that $D(\phi,\eta)$ has the same value as $D(\pi)$ defined earlier in Eq.~\eqref{def:D}, except that we now express it in terms of the new variables $(\phi,\eta)$.  This turns $\mathcal{L}^\dag_{(0)}$ and $\mathcal{L}^\dag_{(1)}$ into
\begin{equation} 
\mathcal{L}^\dag_{(0)}=-\beta\left(1+D\right)\frac{\partial}{\partial\phi},\qquad \mathcal{L}^\dag_{(1)}=\kappa\beta^2\left(1+D\right)^2\frac{\partial^2}{\partial\phi^2},
\end{equation}
where the partial derivatives $\partial/\partial\phi$ are now taken at constant $\eta$.  Thus, the equation that must be solved recursively for higher $i$ terms, takes the form
\begin{equation}
\left(\frac{\partial}{\partial\phi}-\frac{it}{\beta\left(1+D\right)}\right)\chi_{\left(i\right)}=-\kappa\beta\left(1+D\right)\frac{\partial^2}{\partial\phi^2}\chi_{\left(i-1\right)}.
\end{equation}
By introducing the integration factor $\exp\left[-\frac{it}{\beta}\int^{\phi}{\frac{d\phi'}{1+D\left(\eta,\phi'\right)}}\right]=D^{-it/3}$, the solution to this differential equation is easily found to be
\begin{equation}\label{formal solution}
\chi_{\left(i\right)}\left(\phi, \eta;t\right) = \kappa\beta\chi_{\left(0\right)}\left(\phi,\eta;t\right)\int_{\phie}^{\phi}{\frac{1+D\left(\phi',\eta;t\right)}{\chi_{\left(0\right)}\left(\phi',\eta;t\right)}\frac{\partial^2}{\partial\phi'^2}\chi_{\left(i-1\right)}d\phi'}.
\end{equation}
This gives the formal solutions for the different terms of the perturbative expansion. These formal solutions satisfy the correct boundary conditions that completely specify the system.  If we were able to systematically perform all the integrals in~\eqref{formal solution}, we would get the full solution for~\eqref{adjoint Fokker-Planck equation} we wish to solve. 

 Passing from $\chi_{\left(0\right)}$ solution~\eqref{chi-0-solution} to higher $i$'s, it is straightforward to see that $\chi_{\left(1\right)}$ is given by  
 \begin{equation}
\chi_{\left(1\right)}\left(\phi,\eta;t\right)=it\kappa\chi_{\left(0\right)}\left[\ln{\frac{D^{it/3}}{1+D}}-\ln{\frac{W^{it/3}}{1+W}}\right]=it\kappa\chi_{\left(0\right)}\ln\left(\frac{1+W}{1+D}\chi_{\left(0\right)}\right),
\label{perturbative chi1}\end{equation}
where $W$ (without an argument) is meant to be the Lambert function $W(Z)$ of the $Z$ defined in Eq.~\eqref{def:Z-eta}, here and in the sequel.
One can proceed in the same spirit with $\chi_{\left(2\right)}$:
\begin{equation}\label{perturbative chi2}
\begin{aligned}
\chi_{\left(2\right)}\left(\phi,\eta;t\right)=3it\kappa^2\chi_{\left(0\right)}&\left[\frac{2}{1+D}-\frac{2}{1+W}+\left(1-it\right)\ln\left(\frac{1+D}{1+W}\right)\right.\\
&+\left.\frac{2it}{3}\ln \chi_{\left(0\right)}+\frac{it}{6}\ln^2\left(\frac{1+W}{1+D}\chi_{\left(0\right)}\right)\right].
\end{aligned}
\end{equation}
One can check that the normalization condition $\chi\left(t=0\right)=1$ is satisfied at this level of approximation. With the characteristic function at hand, we find the following expression for the mean number of $e$-folds,
\begin{equation}
\left\langle \mathcal{N}\right\rangle=\frac{1}{3}\ln{\frac{D}{W}}-\kappa\ln{\frac{1+D}{1+W}}+O\left(\kappa^2\right)
\label{Naverage}.
\end{equation}
Again, we evaluate the result at $\Phi=\Phi_0$ to obtain the relevant result for the modes that exited the horizon much before the slope change in potential.  It is interesting to note that ${\frac{1+D_0}{1+W_0}}>1$, and consequently in the small diffusion regime, the first crossing time, $\left\langle \mathcal{N}\right\rangle$, is reduced compared to the leading order. In the next section, we will verify that this perturbative result is consistent with the exact result found for the $\beta=0$ case with arbitrary $\kappa$.

One can also calculate the power spectrum
\begin{equation}
  \mathcal{P}_{\mathcal{R}}=\frac{d\langle \delta \mathcal{N}^{2}\rangle}{d\langle\mathcal{N}\rangle},
\label{power spectrum formula}\end{equation}
where $\left\langle \delta \mathcal{N}^{2}\right\rangle$ is found to be
\begin{equation}
\left\langle \delta \mathcal{N}^{2} \right\rangle=\frac{2}{3}\kappa \ln\frac{D}{W}-6\kappa^2\ln\frac{1+D}{1+W}+O\left(\kappa^3 \right).
\end{equation}
The leading contribution to $\mathcal{P}_{\mathcal{R}}$ arising at large scales from quantum diffusion is thus in the form of
\begin{equation}
\mathcal{P}_{\mathcal{R}} =2\kappa\left(1-\frac{6D_0}{1+D_0}\kappa + O(\kappa^2) \right) \approx 2\kappa + 2\varepsilon_2\left( N_0\right) \kappa^2.
 \label{power spectrum31}
\end{equation}
Note that to leading order, $\mathcal{P}_{\mathcal{R}} =2\kappa = (H/2\pi\alpha)^2$, which is consistent with the standard power spectrum of the Starobinsky model we have quoted earlier in Eq.~\eqref{standard-power-limits}.  The next-to-leading term in this equation provides, for the first time, the correction to the power spectrum after a transition using the stochastic $\delta \cal N$ formalism and including the noises of both $\phi$ and $\pi$. This shows that quantum diffusion arising from the small scales yields corrections to scale invariant power spectra that are proportional to the second Hubble parameter at transition time.  Of course, there is also another known contribution to the power of the order of the slow-roll parameter that we have not included here.

\section{The USR Phase After Transition: Nonperturbative Analysis} \label{sec:nonperturbative}

As we noted earlier, for models featuring large noise amplitudes, it is important to study their back-reaction on the large scales to see whether those are immune to physics at small scales or not. We observed that  taking the diffusion matrix similar to the slow-roll case (that is, $\Sigma_{AB}=\left(H/2\pi\right)^2\delta_{A}^{\phi}\delta_{B}^{\phi}$) is not allowed. In fact, the noise in $\pi$ direction is by no means negligible compared to that of $\phi$; especially in the $\beta=0$ case, we have $\Xi_{\pi\pi}=9\Xi_{\phi\phi}$ as long as the condition $\lambda\ll 1$ applies.  This motivates us to study how these noises affect the large scale perturbations in inflationary models with flat potential (or with the so-called ``USR well''~\cite{Pattison:2021oen}). With regard to the Fokker-Planck equation, the noise amplitudes drop exponentially with the same rate, so the associated terms should be considered on an equal footing throughout the USR regime.
 
The phase space transformation introduced in the previous section turns Eq.~\eqref{adjoint Fokker-Planck equation} into
\begin{equation}
\kappa\beta^2\left(1+D\right)^{2}\frac{\partial^{2}\chi}{\partial\phi^{2}}-\beta\left(1+D\right)\frac{\partial\chi}{\partial\phi}+3\beta\kappa\frac{\partial\chi}{\partial\eta}+it\chi=0.
\end{equation}
The $\beta=0$ case corresponds to a USR phase and is our focus in this section. For this, the above relation reduces to
\begin{equation}\label{characteristic function (beta=0)}
3\kappa\left(\eta-\phi\right)^2\frac{\partial^{2}\chi}{\partial\phi^{2}}+\left(\eta-\phi\right)\frac{\partial\chi}{\partial\phi}+\frac{it}{3}\chi=0,
\end{equation}
which may be solved by substituting $\Gamma\equiv\ln\left(\eta-\phi\right)$. We then have
\begin{equation}
\kappa\frac{\partial^{2}\chi}{\partial\Gamma^{2}}-\left(\kappa+\frac{1}{3}\right)\frac{\partial\chi}{\partial\Gamma}+\frac{it}{9}\chi=0.\label{chi(beta=0)}
\end{equation}
This linear equation has a straightforward solution, $\chi=C_{1}\left(\eta\right)\Gamma^{q_1}+C_{2}\left(\eta\right)\Gamma^{q_2}$, where $q_1$ and $q_2$ are given by
\begin{equation}
q_{1,2} = \frac{1}{2} \left( 1 + \frac{1}{3\kappa} \right) \left( 1 \pm \sqrt{1 - \frac{4i\kappa t}{(1 + 3\kappa)^2}} \right).
\end{equation}
The coefficients $C_{1,2}$ could be found by imposing the boundary conditions~\eqref{chi b.c.}, where the UV boundary condition takes the simple form $\frac{\partial\chi}{\partial\phi}\big|_{\phi=\phi_0}=0$. The corresponding solution to Eq.~\eqref{chi(beta=0)} which also satisfies the boundary conditions is easily shown to be
\begin{equation}\label{chi-ab}
\chi = \frac{ q_1 (b/a)^{q_2} - q_2 (b/a)^{q_1} }{ q_1 b^{q_2} - q_2 b^{q_1}}.
\end{equation}
Here 
\begin{equation}
a\equiv\frac{\phie-\eta}{\phi-\eta}, \qquad \text{and} \qquad b\equiv\frac{\phie-\eta}{\phi_{0}-\eta},
\end{equation}
where $a$ determines the starting point $\phi$, and $b$ specifies the relation between the two boundaries; they satisfy $b\leq a\leq1$.  Note also that $\eta = \phi_0 + \pi_0/3 = \phi_{\rm min}$, hence $b=0$ corresponds to setting the left boundary $\phie$ at the farthest classically reachable point $\phi_{\rm min}$.  As a side remark, we consider some limiting cases:  When $t=0$ we have $q_2=0$, so $\chi=1$, as required by the normalizability of $P_\N$.  On the other hand, on the $\pi=0$ axis, the Langevin equation becomes $it\chi=0$; so $\chi$ goes to zero for any $t\neq0$.  Finally, as $\kappa\rightarrow 0$, the exponents $q_1$ and $q_2$ tend to $\infty$ and $it/3$, respectively.  Therefore one finds $\chi\to a^{-it/3}=\chi_{(0)}$, in accordance with~Eq.~\eqref{chi0-beta0}.

\begin{figure}
\centering
\includegraphics[scale=.8]{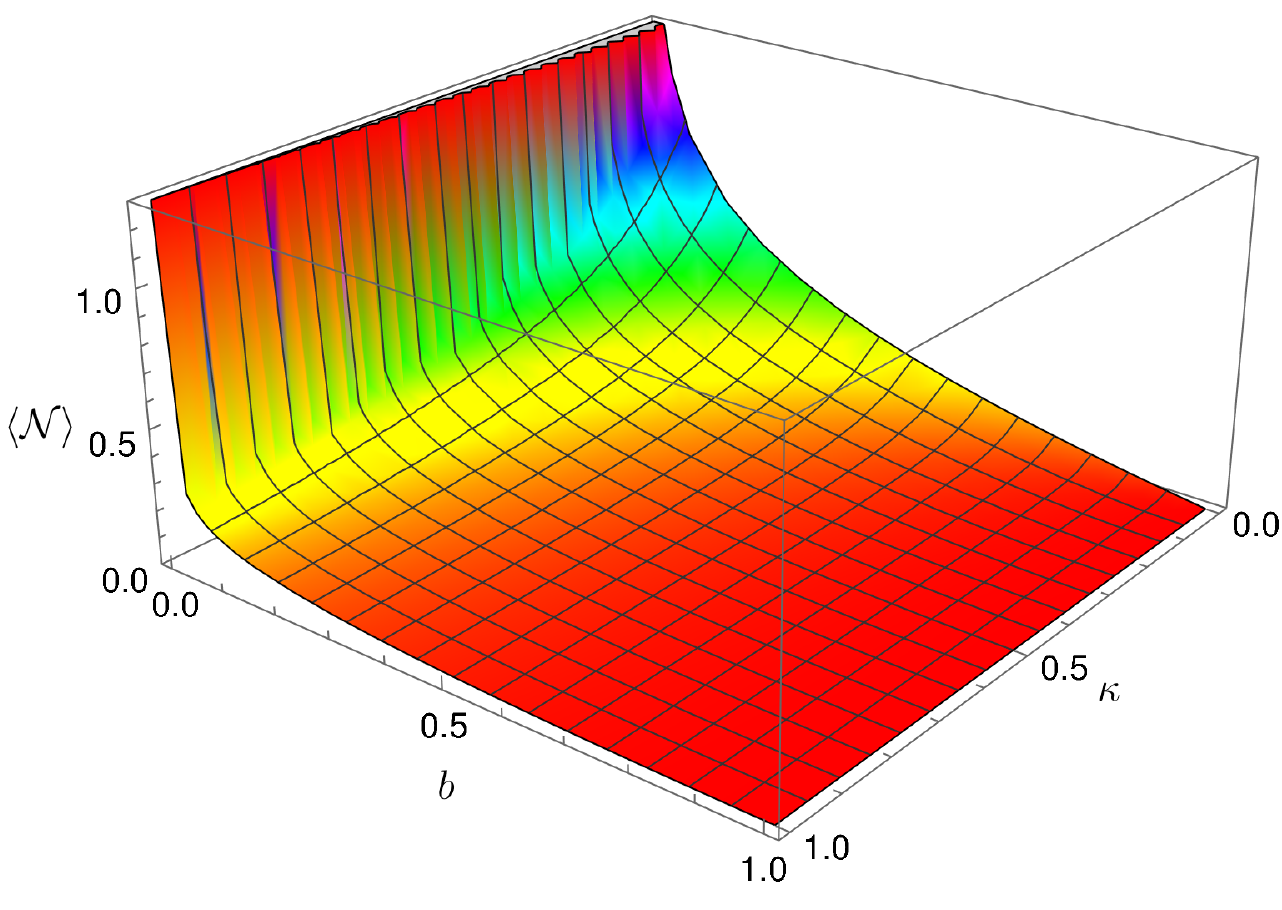}
\caption{Mean number of $e$-folds $\left\langle \mathcal{N}\right\rangle$, calculated from the exact solution, to reach $\phie$ starting from $\phi_0$ under the effect of the noise in phase II.  Note the vanishing of $\left\langle \mathcal{N}\right\rangle$ at $b=1$, and its divergence as $b\to0$.}\label{FiG.Naverage}
\end{figure}

The solution~\eqref{chi(beta=0)} allows one to compute the mean number of $e$-folds and higher order moments of the PDF using Eq.~\eqref{moments}. The complete expression for $\left\langle \mathcal{N}\right\rangle$ is rather cumbersome, but it is straightforward to obtain the relevant result for large scales of our interest by choosing $a=b$,
\begin{equation}\label{mean number exact}
\left\langle \mathcal{N}\right\rangle=\frac{|\ln b|}{3(1+3\kappa)} - \frac{\kappa}{\left(1+3\kappa\right)^2}\left(1-b^{\frac{1+3\kappa}{3\kappa}}\right).
\end{equation}
We have plotted this result for different values of $\kappa$ and $b$ in Figure~\ref{FiG.Naverage}.  Two decaying directions can be identified in $\left\langle \mathcal{N}\right\rangle$, corresponding to increasing either $\kappa$ or $b$:  1) One can see that the contribution of the large diffusion (large $\kappa$) to $\left\langle \mathcal{N}\right\rangle$ is to reduce the mean number of $e$-folds. Furthermore, it shows that not only the classical but also the stochastic phase is short lived.  2)  When $b=1$, Eq.~\eqref{mean number exact} implies $\langle{\cal N}\rangle=0$.  This case corresponds to $\phi_0=\phie$, i.e., the right and left boundaries coincide which means that phase II has vanishing duration and hence does not exist.  At the other extreme, we have $b=0$ where $\langle{\cal N}\rangle$ diverges.  This is the case when the left boundary $\phie$ is equal to $\phi_{\rm min}$, where the classical velocity $\pi$ vanishes.  One may expect that diffusion helps the inflaton reach and get past such classically unreachable points, as is the case in Refs.~\cite{Firouzjahi:2018vet, Pattison:2021oen} where the noise is a constant ($H/2\pi$).  But in our case, the amplitude of the noise ($\propto\pi^2$) happens to vanish exactly at the boundary, so diffusion cannot render $\langle\cal N\rangle$ finite.

Another result which can be obtained from the characteristic function~\eqref{chi-ab}, is the power spectrum, which is found from the moments $\langle{\cal N}\rangle$ and $\langle{\cal N}^2\rangle$ via Eq.~\eqref{power spectrum formula}. Noting that $\chi$ satisfies the Neumann boundary condition $\frac{\partial\chi}{\partial\phi}\big|_{\phi=\phi_0}=0$, it follows by taking successive derivatives with respect to $t$, that the associated moments also satisfy $\frac{\partial\langle{\cal N}^k\rangle}{\partial\phi}\big|_{\phi=\phi_0}=0$. Therefore, in calculating the power for $k<k_0$, one has to apply L'H\^opital's rule, which yields 
\begin{equation}
{\cal P_R} = \lim_{a\to b} \frac{d\langle \delta \mathcal{N}^{2}\rangle}{d\langle\mathcal{N}\rangle} =
\lim_{a\to b} \frac{2\kappa}{\left(1+3 \kappa\right)^2}\frac{2 \left(1 + 6 \kappa \right) x \ln x + \left( 9 \kappa x + 2x + 3 \kappa \right) (x-1)}{\left(1 + 6 \kappa \right)bx - 3\kappa} = 0,
\end{equation}
with $x\equiv\left(b/a\right)^{\frac{1}{3}\frac{1+3\kappa}{\kappa}}$.  It should be noted that this vanishing of the power is a direct consequence of imposing our Neumann boundary condition at $\phi_{\rm UV}=\phi_0$; a more realistic choice would correspond to moving $\phi_{\rm UV}$ far away to the right of $\phi_0$, but we will not do so in the toy model of this section.  Note, however, that the perturbative results of Section~\ref{sec:perturbative} are not affected by the choice of $\phi_{\rm UV}$, since they were based on solving a first-order PDE with only one boundary condition at $\phi_e$.



To describe the tail behavior of the PDF, it is convenient to identify the poles of the characteristic function~\eqref{chi-ab}, as explained in Section~\ref{sec:review-stochastic}. The vanishing of the denominator in Eq.~\eqref{chi-ab} leads to the poles of $\chi$:
\begin{equation}\label{chi-pole-cond1}
b^{q_2 - q_1} = \frac{q_2}{q_1}.
\end{equation}
Let us define some new variables to rewrite Eq.~\eqref{chi-pole-cond1} in a more tractable form:
\begin{equation}\label{def:z,c}
z = \sqrt{1 - \frac{4i\kappa t}{(1+3\kappa)^2}}, \qquad c = -\left( 1 + \frac{1}{3\kappa} \right) \log b > -\log b > 0.
\end{equation}
Note that for a fixed $b$, the $\kappa\to0$ limit corresponds to $c\to\infty$, and $\kappa\to\infty$ corresponds to $c\to-\log b$.  In terms of these variables, Eq.~\eqref{chi-pole-cond1} reads
\begin{equation}\label{chi-pole-cond2}
e^{cz} = \frac{1-z}{1+z}.
\end{equation}
Note that $t$ appears in $z$ while $c$ is independent of $t$, so we can obtain the poles of $\chi$ by solving this equation for $z$. However, it is easy to see that the trivial solution $z=0$ corresponds to $q_1=q_2$ for which both the numerator and the denominator of \eqref{chi-ab} vanish; so $z=0$ does not yield a pole.  We are therefore interested in the nonzero solutions of Eq.~\eqref{chi-pole-cond2}.  It is easy to check that this equation has no real solution except $z=0$.  However, there are infinitely many purely imaginary solutions $z=i\zeta$ with $\zeta\in\mathbb{R}$.\footnote{Inspection shows there are no other complex solutions for $z$.}  In fact, since we are interested in the smallest $it$, it is only the smallest nonzero $\zeta$ that matters (note that the branch cut of the square root in Eq.~\eqref{def:z,c} is chosen such that $\zeta>0$).  Once this smallest $\zeta$ is found, we can plug it back in Eq.~\eqref{def:z,c} and solve for $it$ to obtain the decay rate
\begin{equation}\label{Lambda0}
\Lambda_0 = \frac{(1+3\kappa)^2}{4\kappa} (1+\zeta^2).
\end{equation}
Recall from Eq.~\eqref{chi-pole-cond2} that $\zeta$ itself depends on $\kappa$ and $b$ through $c$.  It is easy to see that, over the range $0<b<1$, $\zeta$ is minimized in the limit $b\to0$ which corresponds to $\zeta\to0$; so for each value of $\kappa$ the minimum decay rate is given by
\begin{equation}\label{Lambda>}
\Lambda_0 \geq \frac{(1+3\kappa)^2}{4\kappa}.
\end{equation}
Interestingly this equation sets a lower bound on $\Lambda_0$, namely,
\begin{equation}\label{Lambda>3}
\Lambda_0 \geq 3.
\end{equation}
This is valid for any value of $\kappa$ and $b$ and is thus a  universal bound.
\begin{figure}
\centering
\includegraphics[scale=1]{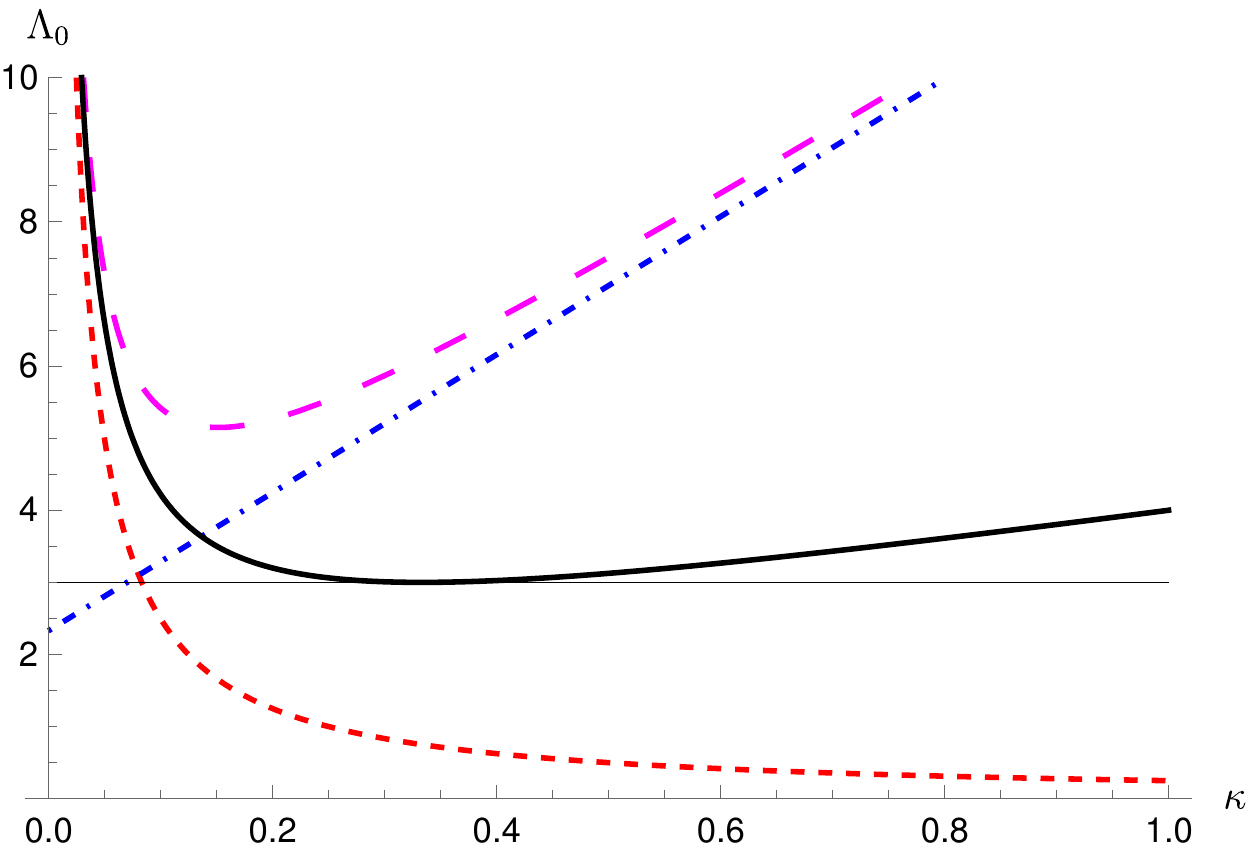}
\caption{The decay rate $\Lambda_0$ as a function of the diffusion strength $\kappa$.  The long-dashed magenta curve is found by solving Eq.~\eqref{chi-pole-cond2} numerically.  The dashed red curve is based on Eq.~\eqref{Lambda0-kappa<<1} and the dot-dashed blue one is based on Eq.~\eqref{Lambda0-kappa>>1}.  These three curves are obtained for $b=0.1$.  The solid black curve is the minimum of $\Lambda_0$ which is attained for $b=0$.}
\label{fig:Lambda0}
\end{figure}

We can obtain approximations for $\Lambda_0$ in the small and large diffusion limits.  Let us first consider the $\kappa\to0$ limit, or more accurately $\kappa\ll\min(1,-\log b)$, which implies $c\to\infty$.  We find the solution to Eq.~\eqref{chi-pole-cond2} in the small $\zeta$ limit (solutions with greater $\zeta$ are irrelevant as we want the smallest pole), which is reduced to $e^{ic\zeta} = 1$, i.e., $\zeta = 2n\pi/c$.  The smallest nonzero $\zeta$ is thus $\zeta = 2\pi/c$.  So we find from Eq.~\eqref{Lambda0} that
\begin{equation}\label{Lambda0-kappa<<1}
\Lambda_0 \approx \frac1{4\kappa} \left[ 1 + O(\kappa) \right], \qquad [\kappa\ll1].
\end{equation}
As can be seen on Figure~\ref{fig:Lambda0}, this approximation agrees with the full numeric result in the $\kappa\to0$ limit.  It is worth mentioning that as $\kappa\to0$, the tail becomes infinitely suppressed, consistent with the Dirac delta fall-off of the PDF in Eq.~\eqref{PDF-delta}.

Now consider the opposite limit $\kappa\to\infty$, i.e., $c\to-\log b$.  This time Eq.~\eqref{chi-pole-cond2} reduces to $b^{-z} = (1-z)/(1+z)$, which has a $\kappa$-independent (but $b$-dependent) smallest solution that we call $z_b=i\zeta_b$.  Therefore, the leading order answer is
\begin{equation}
\Lambda_0 \approx \frac94 (1+\zeta_b^2) \left[ \kappa + O(1) \right],
\end{equation}
which gives the linear approximation with the right slope on Figure~\ref{fig:Lambda0}.  This linear dependence of $\Lambda_0$ on large $\kappa$ is also observed in different models like Ref.~\cite{Pattison:2021oen}.  To obtain the correct intercept, we need to find the $O(1)$ correction too, which includes the $O(1/\kappa)$ correction to $\zeta_b$.  The result is
\begin{equation}\label{Lambda0-kappa>>1}
\Lambda_0 \approx \frac94 (1+\zeta_b^2) \left[ \kappa + \frac23 \frac{2 - \log b}{2 - (1+\zeta_b^2) \log b} + O \left( \frac1{\kappa}\right) \right], \qquad [\kappa\gg1],
\end{equation}
which is in perfect agreement with the full numerical result of Figure~\ref{fig:Lambda0}.

A final limiting case of interest is $b\to0$.  We have already seen that this corresponds to $\langle {\cal N} \rangle \to \infty$, which is inconsistent with exponential tail $\exp(-\Lambda_0 N)$ unless $\Lambda_0\to0$.  In fact, numerical inspection of $\chi$ in the complex $t$ plane reveals that as $b\to0$, the spacing between the poles shrinks to zero (while $\Lambda_0$ remains non-zero), which may also be inferred from the relation $\zeta=2n\pi/c$ that we obtained above in the limit $c\to\infty$.  This amounts to turning a multitude of poles into a branch cut emanating from the branch point $t=-i\Lambda_0$.  Upon Fourier transformation, poles in $\chi$ lead to an exponentially decaying $P_{\cal N}$, whereas a branch cut leads to power-law decay which is heavier and can have infinite $\langle {\cal N} \rangle$.  PDFs with diverging moments are not unheard of in studies of PBHs; they have appeared elsewhere, e.g., in Ref.~\cite{Pattison:2021oen} which has a different noise structure than ours, and in Ref.~\cite{Hooshangi:2021ubn} which is not based on stochastic $\delta\cal N$ at all.


As we remarked earlier our derivation of the noise~\eqref{noise-IIB} was based on the assumption that the noise is not very large.  Therefore, although we have an exact solution for the characteristic function, we cannot claim that it reliably describes the dynamics of the inflaton for large $\kappa$.  Therefore the right branch in Figure~\ref{fig:Lambda0} cannot be trusted.
\section{Conclusions}\label{sec:conclusions}

The statistics of cosmological perturbations in different regimes of the inflationary era is analyzed in the literature with two objectives: 1) finding the back-reaction of quantum diffusion on large scales and 2) providing a phenomenological explanation for a putative population of the PBHs. 

Using the approach of stochastic inflation, we studied the implications of the statistics of coarse-grained cosmological perturbations when a sharp transition occurs in the inflationary potential. Particularly, we focused on perturbations produced in the Starobinsky model, which assumes a linear potential before and after this transition. In this model a slow-roll era (which is characterized by an attractor solution) is followed by a non-slow-roll evolution, in which the second slow-roll parameter $\varepsilon_2$ sharply drops to a negative value. 
This era is unstable and a second slow-roll regime starts after a few $e$-folds.

Comparing with previous studies, we pay special attention to the transition period from the conventional slow-roll to non-slow-roll phase.  In order to study the quantum diffusion beyond slow-roll, we used the stochastic $\delta \mathcal{N}$ formalism. In this context, we focused on the distribution of the first exit time out of the non-slow-roll regime and solved the characteristic function related to that distribution. The back-reaction of small wavelengths on the background dynamics in the framework of stochastic inflation has also been addressed in some previous studies. The Langevin equations governing the coarse-grained phase space variables are usually treated as in a slow-roll regime, i.e, the momentum noise is simply neglected and the field noise amplitude is taken to be $H/2\pi$. Our calculation showed that in a couple of $e$-folds just after a sharp transition, the field and momentum noises should both be taken into account. 
 We argued that this behavior is a common characteristic that may be found in abrupt transitions from slow-roll to another slow-roll regime, with an inevitable non-slow-roll phase in between.

In the low diffusion limit, where quantum diffusion provides a small contribution to the classical trajectories of the Langevin equation, we used a systematic expansion in a small parameter $\kappa$ that characterizes the noise-to-drift ratio. We applied the expansion to the Starobinsky model and showed that if the proper diffusion matrix is used, stochastic effects reduce the mean number of exit time, $\left\langle \mathcal{N}\right\rangle$, out of the transition phase. We also found a correction of order $\varepsilon_2 \mathcal{P}^2_{\mathcal{R}}$ to the power spectrum calculated in standard perturbation theory.  Although the existence of a non-slow-roll phase greatly enhances the perturbations at the transition scale, the presence of the noise due to a sharp transition slightly reduces the power spectrum at large scales.  The difference between the power spectrum resulting from the stochastic inflation and the standard perturbation theory beyond the slow-roll regime was also studied in Ref.~\cite{Ballesteros:2020sre}; they used linear relationship between $\cal R$ and field perturbations and concluded that $\cal P_R$ is unchanged under the inclusion of diffusion. Our analysis showed that there is a small but nonzero contribution by diffusion. 
Furthermore, we showed that in drift-dominated regions quantum diffusion helps to end inflation sooner. We believe that our analysis using $\delta \mathcal{N}$ formalism includes non-linear corrections and is therefore more accurate and reliable.

We also studied the transition from slow-roll to a flat potential (the USR regime) and found an exact expression for the characteristic function. This allowed us to compute the mean number of exit time and higher order moments of the PDF for different realizations, as well as the exact exponential decay rate of the PDF for different values of the noise-to-drift ratio within a toy model. We found that diffusion generally reduces the mean number of exit time. However, in a purely diffusion-driven region inside a flat potential, the probability to end inflation is zero. In other words, in a USR phase after a transition it is impossible for stochastic effects (diffusion) to  dominate, because the velocity and noise amplitude decay in time with the same rate. This result does not coincide with that of figure~2 of Ref.~\cite{Pattison:2021oen}, because the noise used in that work is not induced by a sharp transition.

By tracing these effects on large scale perturbations, one can constrain the inflationary potential models from the time the scale is generated down to the end of inflation. In the Starobinsky model, we identified two more phases of evolution with elaborate form of noises, that one should contemplate on. The effect of quantum diffusion in these regions of potential can also be observed on large scale modes although that might be very small.  This may refine the present results, which we plan to address in a future work.

In our model, the field excursion is limited by a reflecting boundary condition ($\phi_{\rm UV}$) to the post-transition values, once it experienced the USR regime. This assumption and the location of $\phi_{\rm UV}$ can potentially affect our results, although we believe that the above results are chiefly related to the form of noise at the transition instant and beyond.  We also found that in a transition to USR, the power spectrum of the large modes $\left(k<k_0\right)$ vanishes.  This latter result, however, is eminently associated with our particular location of $\phi_{\rm UV}$.

Regarding our second aim, it is well known that the PBHs require large fluctuations to form and a standard perturbation theory approach confirms a local growth of quantum fluctuations in $\mathcal{P}_{\mathcal{R}}$ for modes exiting the horizon during the non-slow-roll phase. Since a perturbative description may not be sufficient, one thus requires a non-perturbative approach such as the $\delta \mathcal{N}$ formalism which predicts exponential rather than Gaussian tail for the PDF~\cite{Ezquiaga:2019ftu}. We found the exponential decay rate for the exit time distribution, by two (mathematically) equivalent methods: 1) analyzing the poles of characteristic function, and 2) directly from the PDF. This rate depends on the field widths of the USR regime as well as the velocity of the field when entering the USR regime. Intriguingly, this rate has a lower bound equal to 3, as found in Eq.~\eqref{Lambda>3}. This heavy tail behavior shows a non-Gaussian probability distribution for large curvature perturbations, i.e., those which are far from the peak of the distribution.  It also shows that those perturbations are largely affected by the quantum diffusion of the fields (in phase space).  

Let us highlight a point regarding the analysis in this work. We worked on a classical trajectory, starting from an attractor in slow-roll regime, as a proxy for different realizations of the stochastic process. We considered the same initial condition, $\Phi$, at the time when the scale of interest crosses out the Hubble radius, for all realizations. In other words, the quantum diffusion in our work plays role \textit{only} after the scale of interest crossed the horizon scale during inflation and receives contributions from all modes that cross out the coarse-graining scale. This assumption partially relies on the results found in Ref.~\cite{Ando:2020fjm}, that the dependence on the initial conditions drops out for the scales in the presence of a dynamical attractor, and is partially based on Ref.~\cite{Grain:2017dqa} which shows that the classical slow-roll attractor is immune to stochastic effects.

In Ref.~\cite{Grain:2017dqa}, it is emphasized that quantum diffusion and the classical slow-roll flow are aligned in the case of test fields with linear equation of motion. This raises a question that we did not address in this paper: Does the same alignment occur during the relaxation phase? Our calculations for the noise at transition epoch showed that diffusion effects and classical flow decay with the same rate. We, therefore, surmise that if there is any misalignment, it would be negligible, though the answer to this question needs more elaboration. 

The discussion in Appendix~\ref{app:transition-noises} about the noise amplitude in a transition epoch, is quite general. It can be applied to models with a rapid change (either decrease or increase) in the field velocity; so it lays the groundwork for extensions to be carried out in the future. 

\section*{Acknowledgements}

We would like to thank V.~Vennin for discussion.  M.N.\ also thanks H.~Firouzjahi and M.H.~Namjoo.  We acknowledge financial support from the research council of University of Tehran.

\appendix

%
%

\appendix
\section{Noise Amplitude in a Transition Epoch}\label{app:transition-noises}
In this appendix, we show that the amplification in velocity noise amplitude is a common feature in all transitions. For this purpose, we do not use arguments based on slow-roll or USR regimes, nor do we need to specify an explicit potential. Instead, we directly work with the Mukhanov-Sasaki equation~\eqref{Mukhanov-Sasaki} at the transition point. Let a single field inflationary scenario start to deviate from slow-roll inflation suddenly at $\eta=\eta_0$. Following our discussion in Section~\ref {sec:background} and recalling that $z=a\pi=-\frac{\pi}{\eta H}$ and $\pi=\frac{d\phi}{dN}$, we note that during a brief non-slow-roll epoch the function $z\left(\eta\right)$ is not described by a simple profile $z=\frac{\text{constant}}{H\eta}$. As a consequence, assuming that the pre-transition stage is governed by slow-roll ($\pi_0=\rm constant$), one obtains
\begin{equation}
z=\frac{\pi_0-\pi}{H\eta}\Theta\left(\eta-\eta_0\right)-\frac{\pi_0}{H\eta},
\end{equation}
and its derivatives can be expressed in the form of
\begin{eqnarray}
z'&=&-\frac{\eta\pi'+\pi_0-\pi}{H\eta^2}\Theta\left(\eta-\eta_0\right)+\frac{\pi_0}{H\eta^2},\nonumber\\
z''&=&-\frac{\pi'(\eta_0^+)}{H\eta}\delta\left(\eta-\eta_0\right)+\text{``finite terms''},
\end{eqnarray}
where $'=\frac{d}{d\eta}$.  Note that although $\pi'$ is discontinuous across the transition, the $\pi'$ that appears in the above equations is meant to be evaluated at $\eta_0^+$.  At the transition point, $z''/z$ can be described by a Dirac-delta function
\begin{equation}
\frac{z''}{z}= \frac{\pi'(\eta_0^+)}{\pi_0} \delta\left(\eta-\eta_0\right).
\end{equation}
This implies that the matching condition on the modes at the transition are 
\begin{equation}
v_{k}^{-}\left(\eta_0\right)=v_{k}^{+}\left(\eta_0\right),\qquad v_{k}^{'-}\left(\eta_0\right)-v_{k}^{'+}\left(\eta_0\right)=\left[\frac{\pi'v_{k}^{-}}{\pi}\right]_{\eta=\eta_0^{+}}.
\end{equation}
Eq.~\eqref{disc-v} is a concrete realization of this formula, related to the Starobinsky model, and $\hat{f}\equiv\frac{1}{3}\frac{d\ln(\pi)}{d\ln(\eta)}\big|_{\eta=\eta_0^{+}}$ is identified by $f$. Here, $\hat{f}$ is related to the rate of $\pi$ variation at the transition point. The calculations continues in the same line of discussion with that in the main part of the paper and the noises' amplitudes right after the transition turn out to be
\begin{equation}\label{noise-IIIAA}
\Xi\big|_{n=n{_0}^{+}}=\left(\frac{H}{2\pi}\right)^2 \begin{pmatrix}
\left[1-\hat{f}\left(1-e^{-3n}\right)\right]^2 & -3\hat{f}\left[1-\hat{f}\left(1-e^{-3n}\right)\right]e^{-3n}\\
-3\hat{f}\left[1-\hat{f}\left(1-e^{-3n}\right)\right] e^{-3n}& 9\hat{f}^2 e^{-6n}
\end{pmatrix}.
\end{equation}
 Using ~\eqref{KG}, one finds that $\hat{f}=\left(1+\frac{V_{,\phi}}{3H^2 \pi}\right)\big|_{\eta=\eta_0^{+}}$ and the ratio of noise amplitude to drift at $\eta=\eta_0^{+}$ is given $\frac{\Xi_{\pi\pi}}{-3\left(\pi+\frac{V'}{3H^2 \pi}\right)}=\frac{\Xi_{\phi\phi}}{\pi}=\frac{H}{2\pi\pi_0}$. One therefore conclude that at a transition with a discontinuity in $\pi'$, the behavior of all noise amplitude matrices are qualitatively similar to what discussed in this paper.  The same qualitative results are expected when $\pi'$ varies rapidly, but not quite discontinuously.
\section{The Solution Surface of Characteristic Equation in Drift-Dominated Limit}\label{app:characteristics}
In this appendix, we solve the leading order characteristic equation 
\begin{equation}
\left[\pi\frac{\partial}{\partial\phi}+3\beta D\frac{\partial}{\partial\pi}\right]\chi_{\left(0\right)}=-it\chi_{\left(0\right)},\ \ \ \ \chi_{\left(0\right)}\left(\phie,\pi;t\right)=1,\label{chi0-A}
\end{equation}
using the method of characteristics. In this method we look for a surface in phase space called the solution surface of the equation. The normal to this surface is given by the vector field $\left(\frac{\partial\chi_{\left(0\right)}}{\partial\phi},\frac{\partial\chi_{\left(0\right)}}{\partial\pi},-1\right)$ and Eq.~\eqref{chi0-A} requires that this vector field be orthogonal to $V=\left(\pi,3\beta D,-it\chi_{\left(0\right)}\right)$. Thus we should look for the integral surface of the vector field $V$, which in turn is generated by its integral curves.
The integral surface of the vector field $V$ is generated by the solution curves of the associated system
\begin{eqnarray}
\frac{d\phi}{\pi}=\frac{d\pi}{3\beta D}=\frac{d\chi_{\left(0\right)}}{-it\chi_{\left(0\right)}}.
\end{eqnarray}
The first integrals of this system are
\begin{equation}
u_{1}\left(\phi,\pi,\chi\right) = \phi + \frac{\pi}{3}-\frac{\beta}{3}  \ln(\beta D), \qquad
u_{2}\left(\phi,\pi,\chi\right) = \frac{ \chi_{\left(0\right)} }{ (\beta D) ^{\frac{it}{3}}}. \label{u1u2}
\end{equation}
The functions $u_{1}$ and $u_{2}$ are functionally independent in a domain of phase space which contains $\phi=\phie$, as an absorbing boundary curve. To find the integral surface of $V$ containing the curve $\phi=\phie$, we first express the curve in parametric form
\begin{eqnarray}
\centering
\phi=\phie,\qquad \pi =s,\qquad \chi_{\left(0\right)}(\phie,s;t) = 1,
\end{eqnarray}
and then compute
\begin{eqnarray}
\left. U_{1}\equiv u_{1} \right|_{\partial\Omega_e}=\phie+\frac{s}{3}-\frac{\beta}{3}\ln \left(\beta D(s)\right), \qquad 
U_{2}\equiv u_{2}|_{\partial\Omega_e}=\frac{1}{\left( \beta D(s)\right)^{\frac{it}{3}}}\label{U1U2},
\end{eqnarray}
where $D(s)$ is the function $D$ on the boundary curve. The first equation of~\eqref{U1U2} gives us the function $D$ in the form of
\begin{eqnarray}\label{b9}
D(s)=W\left(Z\right),\ \ \ \ Z\equiv\frac{1}{e\beta}e^{-\frac{3}{\beta}\left(U_{1}-\phie\right)}.
\end{eqnarray}
Here $W$ is the Lambert $W$ function and $Z$ is a function of $U_{1}$. After eliminating the parameter $s$, one finds
\begin{eqnarray}
U_{2}=\frac{1}{(\beta D(s))^{\frac{it}{3}}}=\left[\beta W\left(\frac{1}{e\beta} e^{-\frac{3}{\beta}\left(U_{1}-\phie\right)}\right)\right]^{\frac{-it}{3}},
\end{eqnarray}
and the required integral surface is given by
\begin{eqnarray}
\chi_{\left(0\right)}\left(\phi,\pi;t\right)=\left[\frac{W\left(Z\right)}{D}\right]^{\frac{-it}{3}}.
\end{eqnarray}
A similar calculation for a flat potential, $\beta=0$, gives us
\begin{equation}
  \chi_{\left(0\right)}=\left(1+\frac{3\left(\phi-\phie\right)}{\pi}\right)^{\frac{-it}{3}}.
\end{equation}
\section{An Alternative Way of Finding PDF Tail Behavior}\label{app:tail}
There is an equivalent method for determining the tail behavior using the eigenvalue problem. In $\beta=0$ case, the adjoint Fokker-Planck equation for $P_{\cal N}(\phi,\pi;N)$ reduces to
\begin{equation}
\kappa  \pi^2 \left[\frac{\partial^2}{\partial\phi^2} - 6\frac{\partial^2}{\partial\phi \partial\pi} + 9\frac{\partial^2}{\partial\pi^2} \right]P_\mathcal{N} + \pi\left(\frac{\partial}{\partial\phi} - 3 \frac{\partial\chi}{\partial\pi}P_\mathcal{N}\right)=\frac{\partial P_\mathcal{N}}{\partial N}.\label{Adjoint Fokker-Planck(beta=0)}
\end{equation}
A suitable variable that turns~\eqref{Adjoint Fokker-Planck(beta=0)} into its canonical form is $\eta=\phi+\frac{\pi}{3}$ and we have
\begin{equation}
LP_\mathcal{N}=\frac{\partial P_\mathcal{N}}{\partial N}, \qquad  L\equiv 9\kappa  \left(\eta-\phi\right)^2\frac{\partial^2}{\partial\phi^2} + 3\left(\eta-\phi\right)\frac{\partial}{\partial\phi}.\label{Adjoint Fokker-Planck(beta=0)2}
\end{equation}
Note that $L$ is a self-adjoint operator with weight function $w\left(\phi\right)=C\left(\eta-\phi\right)^{-2\left(1+\frac{1}{6\kappa}\right)}$. We can find another operator
\begin{equation}
K=\frac{\partial}{\partial\phi}\left(9\kappa \left(\eta-\phi\right)^2\frac{\partial}{\partial\phi} \right)-\frac{1+6\kappa}{4\kappa},
\end{equation}
which is self-adjoint with unit weight and which has the same spectrum as $L$. If we denote the eigenfunctions of $K$ and $L$ by $\Psi$ and $Z$, respectively, those are related by $\Psi=\sqrt{w}Z$. In other words, we have $\left(K+\Lambda_n\right)\Psi_n=0$ and $\left(L+\Lambda_n\right)Z_n=0$. In general, the solution of~\eqref{Adjoint Fokker-Planck(beta=0)2} can be written in the form
\begin{equation}\label{probability}
P_\mathcal{N}=\frac{1}{\sqrt{w}}\sum_{n=0}^{\infty}a_n\Psi_n\left(\phi\right) e^{-\Lambda_n N}.
\end{equation}
It is easy to see that $\Psi_n=C_1\left(\phi-\eta\right)^{m_1}+C_2\left(\phi-\eta\right)^{m_2}$, where
\begin{equation}
m_{1,2}=-\frac 12\left[1\pm\left(\frac{1+3\kappa}{3\kappa}\right)\sqrt{1-\frac{4\kappa}{\left(1+3\kappa\right)^2}\Lambda_n}\right].
\end{equation}
and this gives us
\begin{equation}
Z_n=\tilde{C}_1\left(\phi-\eta\right)^{\tilde{q}_1}+\tilde{C}_2\left(\phi-\eta\right)^{\tilde{q}_2}, \qquad \tilde{q}_{1,2}=m_{1,2}+\left(1+\frac{1}{6\kappa}\right).
\end{equation}
Here $\tilde{q}$ is the same as $q$, in which $it$ is replaced by $\Lambda$. The boundary conditions~\eqref{P_N b.c.},
\begin{equation}\label{PDF b.c.}
\left. \frac{\partial P_\mathcal{N}}{\partial\phi}\right|_{\phi_0}=0, \qquad P_\mathcal{N}\left(\phi=\phie\right)=0,
\end{equation}
imply that $\tilde{C}_1=-\frac{\tilde{q}_2}{\tilde{q}_1}\left(\phi_0-\eta\right)^{\tilde{q}_2-\tilde{q}_1}$ and 
\begin{equation}
\tilde{C}_2\left[\left(\phie-\eta\right)^{\tilde{q}_2}-\frac{\tilde{q}_2}{\tilde{q}_1}\left(\phi_0-\eta\right)^{\tilde{q}_2-\tilde{q}_1}\left(\phie-\eta\right)^{\tilde{q}_1}\right]=0.
\end{equation}
For nontrivial solutions $\tilde{C}_2\neq 0$, we obtain~\eqref{chi-pole-cond1}, in which $q$ is replaced by $\tilde{q}$. In agreement with the discussion of Section~\ref{sec:nonperturbative}, the $\Lambda_n$'s are all positive and real and the exponential decay rate of the PDF for different values of $\kappa$ can be read from Figure~\ref{fig:Lambda0}.

Equation~\eqref{probability} clearly shows an exponential decay in the distribution function of the coarse grained curvature perturbation on all scales across the spectrum, with the same rate. We would like to emphasize that $\Lambda_n$ does not depend on $\phi$ at which the coarse grained mode exits the Hubble radius, although the amplitude $\Psi_n$ depends on the scale. This shows the universality of the form of the tail on all scales.

 We conclude this appendix by noting that the Dirichlet boundary condition~\eqref{PDF b.c.} corresponds to the absorbing prescription that a realization is no longer counted after crossing the boundary $\phi=\phie$. Furthermore, the probability falls exponentially within the timescale given by $\Lambda_0^{-1}$. In other words, $\Lambda_0^{-1}$ plays the role of inflaton lifetime in USR regime. The real, finite and positive values (found in Figure~\ref{fig:Lambda0}) for $\Lambda_0$ imply that all realizations discussed in this paper reach $\phi=\phie$ in a finite number of $e$-folds. This shows that in Starobinsky model, the USR regime is unstable even in the presence of stochastic noises. We emphasize that $a=b=0$ corresponds to $\Lambda_0^{-1}\to\infty$ and infinite number of $e$-folds required to exit.


\bibliography{stochastic-usr-transition}{}
\bibliographystyle{JHEP.bst}

\end{document}